\documentclass[11pt,onecolumn, draftcls]{IEEEtran}

\usepackage{amsmath, amssymb}

\usepackage{graphicx}

\newtheorem{theorem}{Theorem}
\newtheorem{proposition}{Proposition}

\newtheorem{lemma}{Lemma}
\newtheorem{observation}{Observation}

\newtheorem{corollary}{Corollary}
\newtheorem{definition}{Definition}
\newtheorem{example}{Example}
\newtheorem{problem}{Problem}

\begin{document}

\title{Max-Flow Min-Cut Theorems for Multi-User Communication Networks}

\author{S\o ren Riis and Maximilien Gadouleau \thanks{
Riis is with the School of Electronic Engineering and Computer Science at
    Queen Mary University of London. Gadouleau is with School of Engineering and Computing Sciences at University of Durham.

\noindent Emails: s.riis@qmul.ac.uk and m.r.gadouleau@durham.ac.uk

\noindent
 This work  was supported by EPSRC  ref: EP/H016015/1}}

\maketitle

\begin{abstract}

The paper presents four distinct new ideas and results for communication networks: 

1) We show that relay-networks (i.e. communication networks where different nodes  use the same coding functions) can be used to model dynamic networks, in a way, vaguely akin to Kripke's {\em possible worlds} from logic and philosophy. Link failures, point failures, changes in network topology during transmission, changes in receivers demands etc. can all be modelled, in a discrete fashion, by considering a multiverse where different possible worlds (hypothetical situations) are modelled as worlds existing in parallel. Nodes with the same labels might represent nodes in parallel worlds that behave, in the same way, as they are unaware of which world becomes the actual world.  
 
2) We introduce {\em the term model}, which is a simple, graph-free symbolic approach to communication networks. We use the term model to create an algorithm (based on the max-flow min-cut algorithm) that calculates the information-theoretic limit for the capacity of a given communication network. We notice that different non-isomorphic communication networks might never-the-less be mathematically identical because they lead to the same term model. We illustrate the power of our formalism through a number of examples. 

3) We state and prove variants of a theorem concerning the dispersion of information in single-receiver communications. The dispersion theorem resembles the max-flow min-cut theorem for commodity networks and states that the minimal cut value (the channel capacity) can be achieved asymptotically. The theorem is proved by combining Menger's theorem from graph theory with an argument that is similar to the proof of Shannon fundamental theorem for memoryless noisy channels.   To prove the theorem we introduce a very weak kind of network coding (network coding lite), which we will refer to as routing with dynamic headers.

4)  We show that the solvability of an abstract multi-user communication problem is equivalent to the solvability of a single-target communication in a suitable relay network. 

In the paper, we develop a number of technical ramifications of these ideas and results.  One technical result is a max-flow min-cut theorem for the R\'enyi entropy with order less than one, given that the sources are equiprobably distributed; conversely, we show that the max-flow min-cut theorem fails for the R\'enyi entropy with order greater than one.  We leave the status of the theorem with regards to the ordinary Shannon Entropy measure (R\'enyi entropy of order one and the limit case between validity or failure of the theorem) as an open question.
In non-dynamic static communication networks with a single receiver, a simple application of Menger's theorem shows that the optimal throughput can be achieved without proper use of network coding i.e. just by using ordinary packet-switching.  This fails dramatically in relay networks with a single receiver. We show that even a powerful method like linear network coding fails miserably for relay networks.  With that in mind, it is noticeable that our rather weak form of network coding (routing with dynamic headers) is asymptotically sufficient to reach capacity.

\end{abstract}

\section{Introduction} \label{sec:intro}

There is an extensive literature on commodity networks and these play a central role in an array of applications: traffic routing, urban planning, scheduling of freight delivery,  economical networks, communication through packet switching and much more.  Though a typical commodity network has multiple sources  and multiple sinks, the central theorem in the field - the max-flow min-cut theorem - only concerns the special case of commodity networks with a single source and a single sink. 

Digital information in communication networks is radically different form ordinary commodities as it can be modified, copied or mixed during transmission.  This idea goes beyond traditional routing and lays the foundation of Network Coding \cite{YZ99, ACLY00}. In Network  Coding intermediate nodes can combine the packets they receive and retransmit the combined versions towards their destinations.  An instance of Network Coding is given by a network with prescribed sets of sources and of destinations, where the destinations request messages sent by the sources.  In this paper we generalise network coding networks to relay networks (transfer networks) that are networks where different nodes might be required to use the same coding function.  Our main result is a theorem that mathematically plays a similar role in the theory for relay networks as the max-flow min-cut theorem plays in the theory of commodity networks. 

The general problem of determining whether all the demands of the receiver nodes in a communication network can be satisfied simultaneously has been widely studied \cite{DFZ04, DFZ07} since the discovery of network coding. The problem of solvability of multi-user communications exhibits many pathological examples, for instance there exist communication problems which are asymptotically solvable but not solvable for any finite alphabet \cite{DFZ06}. Different methods have been proposed to determine whether a problem is solvable, including graph entropy \cite{DFZ07} and guessing games \cite{Rii07}.

\subsection{Relay networks for communication networks} \label{sec:intro_relay}

\begin{figure}[!htbp]
\begin{center}
    \includegraphics[scale=0.55]{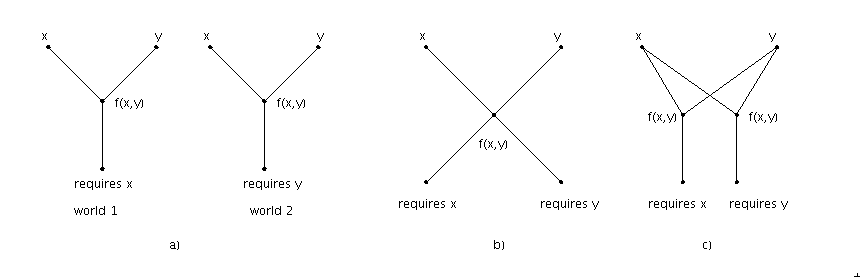}
\end{center}
\caption{The receiver node requires message $x$ or message $y$, but the internal node does not know which is the case. The dynamic communication problem  is equivalent to the relay network in Figure 1b as well as the relay network in Figure 1c. } \label{fig:fig01}
\end{figure}

Let us consider a few basic examples that introduce dynamic communication networks and illustrate their link to relay networks. This will be properly formalised in Sections \ref{sec:NC_terms} and \ref{sec:extension}.

\begin{example}
In Figure 1 we consider a dynamic network with a single receiver. This receiver might require the message $x$ (world 1) or might require the message $y$ (world 2). As usual in the literature on network coding we assume that the messages $x$ and $y$ are selected from some finite message space (finite alphabet $A$), that $f:A \times A \rightarrow A$ is a suitable coding function and that $f(x,y)$ is some message selected from $A$.  And we assume that each edge has unit capacity i.e. allows the transmission of one message (e.g. a block of symbols) \footnote{For a more detailed account see \cite{YZ99} or any other standard text on Network Coding}. The communication problem in Figure 1 is unsolvable in the sense that it is impossible to satisfy the receivers demands (at a full rate where each receiver, receives one message in each step).  The relay networks in Figure 1b and Figure 1c illustrate how the dynamic network can be represented as relay networks.
\end{example}

\begin{figure}[!h]
\begin{center}
    \includegraphics[scale=0.55]{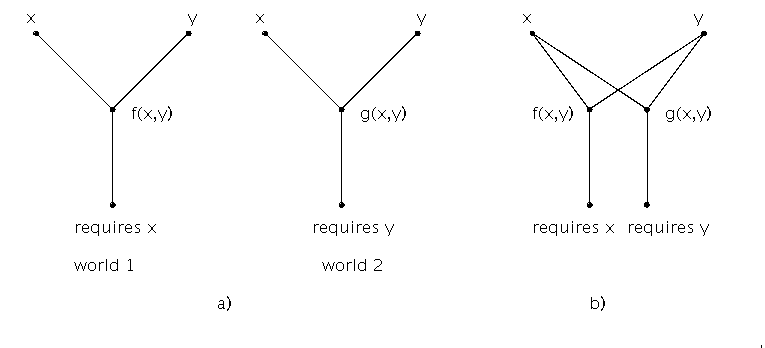}
\end{center}
\caption{The receiver requires message $x$ or message $y$ and the internal node knows which is the case and can choose its coding function accordingly. The dynamic communication problem  is equivalent to the relay network in Figure 2b. } \label{fig:fig02} 
\end{figure}

\begin{example} In Figure 2, we consider the same network, however here we assume that the inner node knows the requirement of the receiver, so it can choose the routing accordingly by either applying the coding function $f:A \times A \rightarrow A$ given by 
$f(x,y)=x$ or by applying the coding function $g:A \times A \rightarrow A$ given by $g(x,y)=y$ depending on the situation. The situation can be represented as a static communication problem with two receivers as seen in Figure 2b.
\end{example}

 \begin{figure}[!htbp]
\begin{center}
    \includegraphics[scale=0.40]{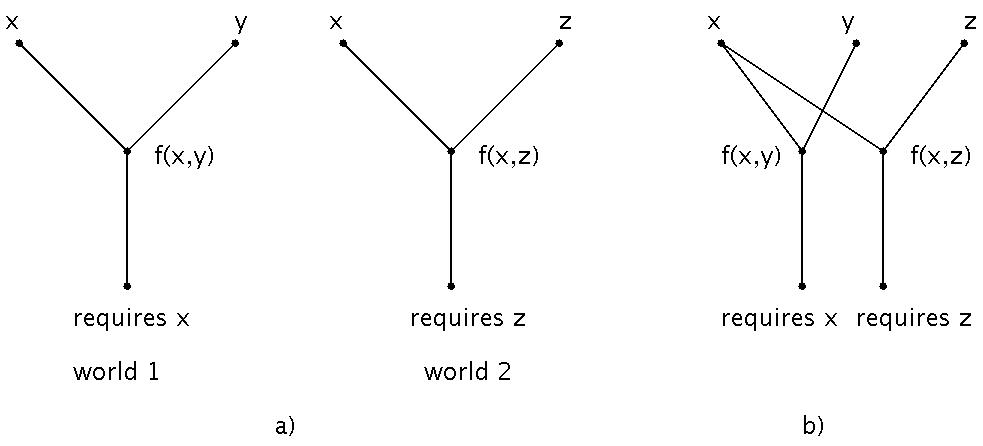}
\end{center}
\caption{The receiver prefers message $z$ to $x$ and is the least interested in message $y$.  The internal node does not know the preferences for the receiver.} \label{fig:fig03}
\end{figure} 

\begin{example}
In Figure 3 the receiver has a preference between the messages so message $z$ is preferred to message $x$ that in turn is preferred to message $y$. 
The sender nodes are aware of this preference, but the inner node is oblivious to the situation and always broadcasts a message determined by a fixed coding function $f: A\times A \rightarrow A$. 
 \end{example} 

\begin{figure}[htb]
\begin{center}
    \includegraphics[scale=0.25]{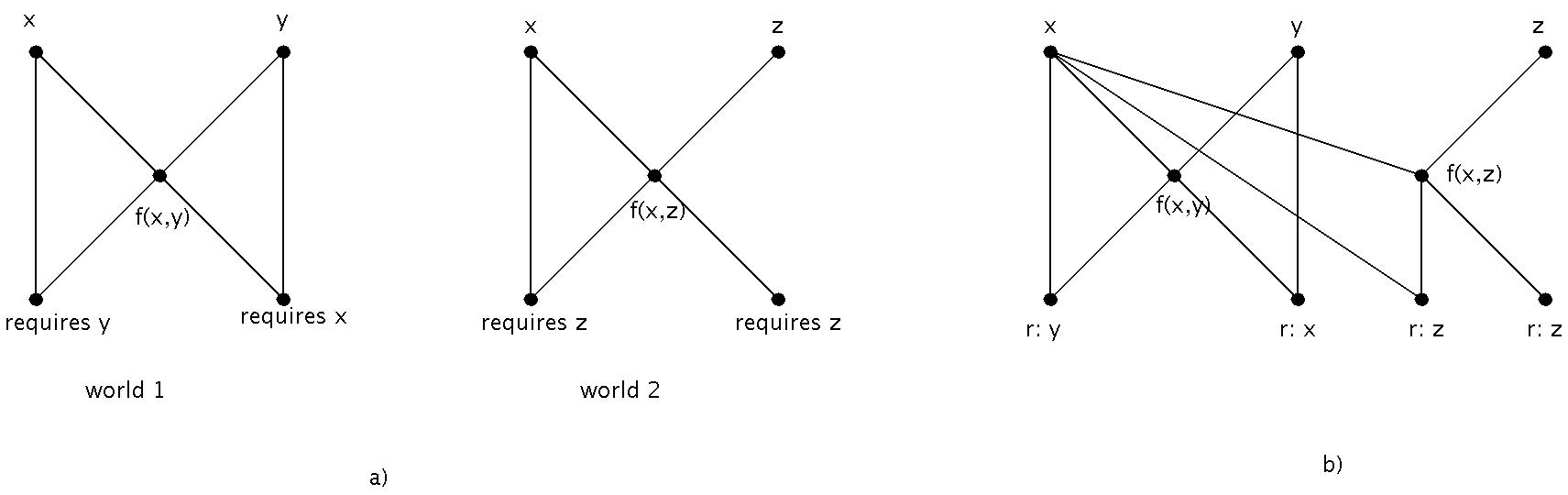}
\end{center}
\caption{The receiver node requires message $x$ or message $y$, but the internal node does not know which is the case. The relay network in b) is equivalent to the dynamic communication problem in a).} \label{fig:fig04}
\end{figure}

\begin{example}\label{exam04}
In Figure 4 there are two receivers. World 1 is the butterfly network that can be solved using network coding. In world 2 there is a link failure, and in conjunction with this link failure an important message $z$ is being broadcast to both receivers. The inner node is not concerned with the situation and uses the same coding function $f:A \times A \rightarrow A$ in both cases.  The communication problem is equivalent to the communication problem given by the relay network with 4 receivers given in Figure 4b. 

\subsection{Term sets and relay networks} \label{sec:intro_term}

Mathematically, the problem in example 3 can be described as the task of constructing a (coding) function $f:A \times A \rightarrow A$ such that $x \in A$ can be reconstructed from the value $f(x,y) \in A$, and $z \in A$ can be reconstructed from the value $f(x,z) \in A$.  Explicitly, we are looking for (decoding) functions $h_1,h_2: A \rightarrow A$ 
such that the {\em term equations} $h_1(f(a_1,a_2))=a_1$ and $h_2(f(a_1,a_2))=a_2$ hold for all $a_1,a_2 \in A$. While this is impossible for finite alphabets, asymptotically for large alphabets $A$ (e.g. by considering long blocks of bits) we can do very well by sending messages $x,y$ and $z$ such that a small {\em header} part of the messages indicate if the {\em body} of message is $x,y$ or $z$ \footnote{In the actual example one bit suffices that is $0$ for messages $x$ and $y$ and is $1$ for message $z$}. The coding function $f$ simply outputs the value of $x$ or $z$ according to this header information.  
\end{example}

In general it turns out that a given communication problem given by a relay network can be expressed by a set of term equations.  This graph free symbolic approach cuts right to the mathematical part of the given communication problem. Non-isomorphic relay networks (e.g. the ones in Figure 1b and Figure 1c) might lead to the same set of term equations. One of the key points in working with term equations rather than the graphs is the flexibility this gives.  We can for example translate the relay problem in Figure 3b that has two receiver nodes, to a problem that - like the dynamic version of it in Figure 3a - has only one receiver. In this problem the receiver has access to the terms $f(x,y)$ and $f(x,z)$ and requires the values of $x$ and $z$.  Or explicitly, the task: 

\bigskip
\noindent
(1) {\em Find functions $f,h_1$ and $h_2$ such that $h_1(f(x,y))=x$ and $h_2(f(x,z))=z$ hold for all $x,y,z \in A$.}

\bigskip

This task is logically equivalent to (if we only are concerned with solvability) 

\bigskip
\noindent
(2)  {\em Find functions $f,h_1$ and $h_2$ such that  $h_1(f(x,y))=x$ and $h_2(f(x,y))=y$ holds for all $x,y \in A$.}

\bigskip

Notice that the communication problem in Figure 1b gives rise to exactly the same task as (2). In general we are interested not just in solvability or unsolvability, but also in how much information that can be transmitted to the receivers. With regard to that type of question, the term equations in (1) and (2) are distinct.  

Similarly, we can translate the problem in Figure 4b - a network communication problem with four receiver nodes - to a problem with only a single receiver with access to the terms $y_1,f(x_1,y_1),x_2,f(x_2,y_2)$ and $f(x_3,z)$ that requires the messages $x_1,y_2$ and $z$.  More specifically the task is to

\bigskip
\noindent
(3) {\em Find functions $h_1,h_2,h_3$ such that $h_1(y_1,f(x_1,y_1))=x_1, h_2(x_2,f(x_2,y_2))=y_2$ and $h_3(f(x_3,z),x_3)=z$ for all $x_1,y_1,x_2,y_2,x_3,z \in A$.}
\bigskip

While task (3) can not be achieved, it can asymptotically be achieved with regards to some measures as we can divide the message space into two parts: One part for messages $x$ and $y$ and one part for messages $z$. If messages are transmitted as blocks of bits,  we can use one of the bits as a {\em flag} that controls the routing at the inner node. The remaining bits can then be used freely to send the body of the messages $x,y$ and $z$. 

In the paper we show that this type of routing - where a negligible part of the message serves as header information - can asymptotically achieve the capacity for single receiver relay networks.  More specifically  we show that the above example works for general relay networks. That is we will consider multi-user communication problems where  each user is assigned a collection of variables (representing the required messages) and as well as a set of  {\em terms} (as known from logic) expressing the relationship between the sent messages, the coding functions and the received messages.  A term is built on variables representing the messages sent by the sources, and on function symbols representing coding functions at the intermediate nodes. A term thus formally represents all the possible operations undergone by messages from the sources to the destinations. This novel representation has several advantages. First, the topology of the network is contained in the term set; we can hence work without the help of the adjacency matrix of the network. Second, this graph-free framework makes computations easier to handle. Third, it is versatile and allows to convert a dynamic multi-user network process into a static  single static system. Fourth, it is actually more general than network coding, and hence offers not only a generalisation of results in network coding, but also a reformulation in terms of flows.

\subsection{Max-flow min-cut for term sets} \label{sec:intro_min-cut}

Next, and this is a crucial for our main results, we define the {\em min-cut} of a term set, which can be viewed as the channel capacity to the receiver
and hence represents the information bottlenecks on the network. Conversely, to each choice of coding functions we associate a flow value, referred to as the {\em dispersion}, which quantifies the amount of information sent to the receivers. More precisely, the dispersion is the logarithm of the number of possible outputs of the term set, while the {\em one-to-one dispersion} is the logarithm of the number of outputs with exactly one pre-image, i.e. for which the input can be completely determined.

In the paper we show a max-flow min-cut theorem for the dispersion of term sets: the maximum dispersion and one-to-one dispersion of a term set are asymptotically equal to the value of its min-cut. 

The term sets we consider may have distributed coding functions, which happens when different subterms use the same coding function. For instance, distributed coding functions occur in the term set associated to a multi-user communication problem, where a distributed function represents the same intermediate node in terms received by different users. Our proof of the max-flow min-cut theorem is based on a novel protocol, referred to as {\em dynamic routing}, which uses dynamic headers to eliminate distributed functions. Clearly, this comes at a cost in bandwidth equal to the size of the header; however, this is a constant given by the term set and becomes negligible when the alphabet size increases. Dynamic routing is interesting in its own sake, for unlike typical network coding approaches, such as random linear network coding \cite{HMK+06}, the manipulation of data is operated on headers only, and not on the whole packets.

If all sources are cooperative and can choose the optimal input distribution, then the maximum amount of information that can be inferred about the input from the received output is given by the min-cut of the term set. We thus introduce different measures of performance based on the R\'enyi entropy \cite{Ren61} for the non-cooperative case where the inputs are uniformly distributed.
We show that the dispersion is a special case of the R\'enyi entropy, while the one-to-one dispersion is an independent performance measure.

The second main contribution is the max-flow min-cut theorem for the R\'enyi entropy with order $0 \leq \alpha < 1$, thus strengthening the result for the dispersion. Conversely, the R\'enyi entropy for $\alpha > 1$ does not necessarily reach the min-cut. Therefore, the R\'enyi entropy is sensitive to information bottlenecks that cannot be taken into account via the min-cut approach. However, the case of the Shannon entropy, where $\alpha = 1$, remains open.

In order to simplify the combinations operated at each intermediate node and the decoding at each destination, linear network coding only considers linear coding functions \cite{LYC03, KM03}. It is known that linear network coding is not optimal in general \cite{Rii04, DFZ05}; we generalize the inefficiency of linear network coding in our framework. In particular, we design a family of term sets with arbitrarily large min-cut where the maximum dispersion achieved by linear functions is only equal to $2$. This can be intuitively explained by the fact that the dispersion of linear coding functions is equal to their min-entropy, which is the R\'enyi entropy of infinite order. Conversely, we prove that if maximum dispersion equal to the min-cut can be achieved using coding and decoding functions based on polynomials of fixed degree, then it can be achieved using linear functions only.

The third main contribution is the multi-user max-flow min-cut theorem. This shows that the maximum dispersion received by each receiver (user) can be asymptotically attained simultaneously. In other words, if a dispersion can be achieved locally, i.e. while disregarding the other users, it can be achieved globally, i.e. when the other users have to be accommodated as well. This result is then applied to multi-user communication problems such as satellite communication (the well-known butterfly network) and data storage.

Finally, our framework based on term sets is extended to simulate dynamic networks whose topologies may change over time. We view a dynamic network as possible ``worlds'', i.e. states in which the network is, and we allow the users to have requirements on the dispersion that change over time. A dynamic network can thus be modeled as one main term set, viewed as the union of all term sets for all users, possible worlds, and time-slots. Our last main contribution is the multi-user theorem for dynamic networks, which proves that if the demand (over all worlds and all time-slots) of each user can be satisfied locally, then they can all be satisfied globally.

\subsection{Outline}

The rest of the paper is organized in two main parts as follows. Firstly, Sections~\ref{sec:comm_N_logic} to \ref{sec:case_study} study term sets and relay networks. Section~\ref{sec:comm_N_logic} reviews some key concepts of logic and term sets and defines the analogues of flows and cuts in the new communication networks. Section~\ref{sec:max-flow_min-cut} then proves the max-flow min-cut theorems for the dispersion and the one-to-one dispersion for these networks. The theorem for the R\'enyi entropy is given in Section \ref{sec:renyi}. Section~\ref{sec:linear} then investigates the dispersion of linear coding functions. In order to illustrate the concepts and results of this paper, a case study of a simple term set is carried out in Section~\ref{sec:case_study}. Secondly, Sections \ref{sec:NC_terms} and \ref{sec:extension} illustrate how relay networks, term sets, and the max-flow min-cut theorems can be applied to communication networks. In Section~\ref{sec:NC_terms}, we associate a term set to a multi-user communication problem, and prove the multi-user max-flow min-cut theorem. Our model is finally generalized to dynamic networks in Section \ref{sec:extension}. Section \ref{sec:conclusion} then concludes summarizes the paper.

\section{Abstract Communication Channel based on logic} \label{sec:comm_N_logic}

This section introduces a new type of abstract communication channels (1-1 user communication networks) based on term sets in logic and determines its main characteristics. We first review the basic concepts of logic and determine the analogue of a min-cut.
We then view flows as transmission of data over a given alphabet, hence determining the analogue of max-flow.

\subsection{Term sets} \label{sec:logic}

Let $X = \{x_1,x_2,\ldots,x_k\}$ be a set of {\em variables} and consider a set of {\em function symbols} $\{f_1,f_2,\ldots,f_l\}$ with respective arities (numbers of arguments) $d_1,d_2,\ldots,d_l$. A {\em term} is defined to be an object obtained from applying function symbols to variables recursively. For instance, if $k=2$, $l=3$, and the arities are given by $d_1 = 1$, $d_2 = d_3 = 2$, then the following are terms: $t_1 = f_2\big(f_1(x_1),x_2\big)$, $t_2 = f_1\Big(f_3\big(f_2(x_2,x_1),f_3(x_1,x_2)\big)\Big)$, $t_3 = f_1(x_1)$.
For a broader introduction to first-order logic see for example \cite{B77}. Only Definition 3.2 on p.18 which introduces the notions of variables, function symbols, and terms is relevant for us.

We say that $u$ is a subterm of $t$ if the term $u$ appears in the definition of $t$. For instance, $t_3$ is a subterm of $t_1$ as $t_1 = f_2(t_3,x_2)$, but it is not a subterm of $t_2$. Furthermore, $u$ is a {\em direct subterm} of $t$ if $t = f_j(v_1,\ldots,u,\ldots,v_{d_j})$, and $f_j$ is referred to as the {\em principal function} of $t$.

We shall consider finite term sets, typically referred to as $\Gamma = \{t_1,t_2,\ldots,t_r\}$ built on variables $x_1,x_2,\ldots,x_k$ and function symbols $f_1,f_2,\ldots,f_l$ of respective arities $d_1,d_2,\ldots,d_l$. We denote the set of variables that occur in terms in $\Gamma$ as $\Gamma_{{\rm var}}$ and the collection of subterms of one or more terms in $\Gamma$ as $\Gamma_{\rm sub}$; thus $\Gamma_{{\rm var}} \subseteq \Gamma_{\rm sub}$ and $\Gamma \subseteq \Gamma_{\rm sub}$.

\begin{definition} \label{def:channel}
A {\em (network coding) communication channel} is given by a collection $\Gamma$ of terms. 
The {\em requirement} of the channel is a collection of variables. If no such set is specified it is assumed that the channel requires all variables.

A {\em communication network} is a collection of receivers  that each is assigned a channel with some requirement. 

A communication network is said to be a {\em many-to-many (or multi-cast) network} if each receiver requires the same set of variables. 
\end{definition}

\begin{example} \label{example:butterfly}
The {\em butterfly network} has two receivers. One receiver is assigned a channel given by $\{x, f(x,y)\}$ and requires $y$. The other receiver is assigned a channel given by $\{y,f(x,y)\}$ and requires $x$.

We rename variables and take the union of terms sets for the butterfly network we get a communication channel 
$\Gamma:= \{x,f(x,y),w,f(z,w)\}$  which requires $y,z$ (or equivalently requires $x,y,z,w$).
\end{example}

We now define a term-cut, which can be viewed as replacing some subterms in the definition of a term by variables.

\begin{definition}[Term-cut] \label{def:term-cut}
A set of subterms $s_1,s_2,\ldots,s_{\rho} \in \Gamma_{\rm sub}$ provides a {\em term-cut} of size $\rho$ for $\Gamma$ if all the terms can be expressed syntactically by applying function symbols to $s_1,s_2,\ldots,s_{\rho}$.

For a collection of terms $\Gamma$ and for a collection of variables $U \subseteq \Gamma_{{\rm var}}$ we let 
$\Gamma^U$ denote the set of terms that occur by substituting each variable $x \in \Gamma_{{\rm Var}} \setminus U$ with the constant symbol $0$.

A set of subterms $s_1,s_2,\ldots,s_{\rho} \in \Gamma_{\rm sub}$ together with the symbol $0$ provides a {\em term-cut with respect to the variables in} $U$  of size $\rho$ for $\Gamma$ if all the terms can be expressed syntactically by applying function symbols to $s_1,s_2,\ldots,s_{\rho}$ and $0$.

\end{definition}

A minimal term-cut for $\Gamma$ is a term-cut with minimum size, referred to as the {\em min-cut} of $\Gamma$. The min-cut can hence be viewed as the number of degrees of freedom of the term set. Alternatively, the min-cut can be viewed as a measure of the channel capacity of the channel defined by $\Gamma$.
Clearly, the min-cut is no more than the number of variables $k$ since $\{x_1,x_2,\ldots,x_k\}$ is a term-cut for $\Gamma$; similarly, the min-cut is no more than the number of terms $r$.

\begin{example} \label{example:term-cut}
Consider the communication channel given by the term set
\[
    \Gamma_1=\left\{h\big(f(x,y), g(z,w), f(y,x)\big), m\big(g(z,w), f(y,x)\big), g\big(f(x,y), g(z,w)\big), f\big(g(z,w), f(y,x)\big) \right\},
\]
then the subterms $s_1 = f(x,y)$, $s_2 = g(z,w)$, and $s_3 = f(y,x)$ form a term-cut for $\Gamma_1$ since we--in a purely syntactical way--can express the terms in $\Gamma_1$ by applying function symbols to $s_1$, $s_2$, and $s_3$ as
\[
    \Gamma_1 = \{h(s_1,s_2,s_3), m(s_2,s_3), g(s_1,s_2), f(s_2,s_3)\}.
\]
This shows that $\Gamma_1$ has capacity $3$ (a simple combinatorial argument shows there is no min-cut of value less than $3$). 

The min-cut of $\Gamma_1$ with regards to the variables $w,z$ is $1$ as

\[
    \Gamma_1^{\{w,z\}} =\left\{h\big(f(0,0), g(z,w), f(0,0)\big), m\big(g(z,w), f(0,0)\big), g\big(f(0,0), g(z,w)\big), f\big(g(z,w), f(0,0)\big) \right\},
\]
has a term cut with $s_1=g(z,w)$ as we can express each term in $\Gamma_1^{\{w,z\}}$ by applying function symbols to $s_1$ and $0$ since

\[
    \Gamma_1^{\{w,z\}} = \{h(f(0,0),s_1,f(0,0)), m(s_1,f(0,0)), g(f(0,0),s_1), f(s_1,f(0,0))\}.
\]

Thus the value of the minimal term cut of $\Gamma_1$ is $1$ with respect the variables $w,z$.

\end{example}

The concepts explained so far can be graphically explained as follows.

\begin{definition}[The graph $G_\Gamma$] \label{defi:G_Gamma}
For a given term set $\Gamma$, the directed graph $G_{\Gamma}=(V,E,S,T)$ is defined to have vertex set $V= \Gamma_{\rm sub}$, edge set $E = \{(u,v) : u \,{\it is\ a\ direct\ subterm\ of}\, v\}$, source set $S=\Gamma_{{\rm var}}$, and target set $T=\Gamma$.
\end{definition}

In the graph $G_\Gamma$, each term is connected to all the variables it is built on; however, the graph is clearly acyclic. Notice that $S \cap T$ is non-empty if $\Gamma$ contains one or more terms that are variables.

\begin{figure}
\begin{center}
    \includegraphics[scale=0.50]{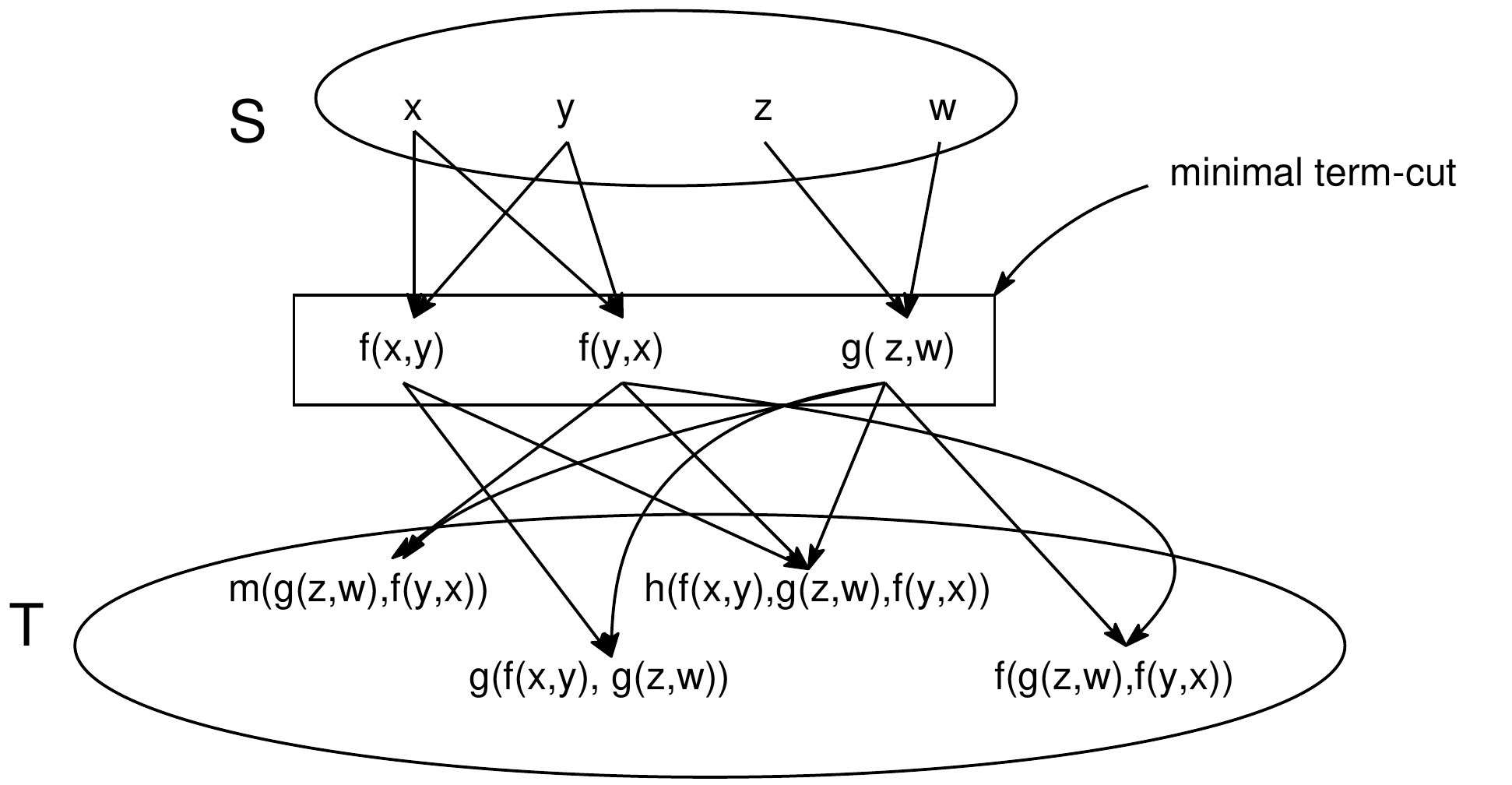}
\end{center}
\caption{The graph $G_{\Gamma_1}$.} \label{fig:termcutgraph}
\end{figure}

\begin{example} \label{example:graph_gamma}
Consider the term set $\Gamma_1$ in Example~\ref{example:term-cut}.
The graph $G_{\Gamma_1}$ consists of a vertex for each subterm in
\begin{eqnarray*}
    \Gamma_{1,{\rm sub}}&=& \left\{x,y,z,w,f(x,y), f(y,x), g(z,w), h\big(f(x,y), g(z,w), f(y,x)\big), m\big(g(z,w), f(y,x)\big),\right.\\
     &&\left. g\big(f(x,y), g(z,w)\big), f\big(g(z,w), f(y,x)\big)\right\}.
\end{eqnarray*}
Furthermore, each variable in $ \Gamma_{1,{\rm var}}=\{x,y,z,w\}$ represents a source node and each term in $\Gamma_1$ represents a sink (or target) node. The graph $G_{\Gamma_1}$ is then given as in Figure~\ref{fig:termcutgraph}.
\end{example}

Assume that $G$ is a directed graph with source set $S$ and target set $T$. We say a set $U$ of vertices  is a {\em vertex cut}---commonly referred to as a separating set---if the removal of $U$ leaves no {\em directed} path from $S$ to $T$. If $S \cap T \neq \emptyset$ each single point in $S \cap T$  is considered to be a path from $S$ to $T$. Proposition~\ref{prop:two_cuts} below shows that term-cuts for $\Gamma$ are equivalent to vertex cuts in $G_\Gamma$.

\begin{proposition} \label{prop:two_cuts}
Assume $\Gamma$ is a finite term set.  A subset $C \subseteq \Gamma_{\rm sub}$ is a term-cut for $\Gamma$ if and only if $C$ is a vertex cut that separates $S=\Gamma_{{\rm var}}$ from $T=\Gamma$ in the directed graph $G_{\Gamma}$. Therefore, the min-cut of $\Gamma$ is identical to the size of the minimal cut that separates $S$ from $T$ in the directed graph $G_{\Gamma}$.
\end{proposition}

The proof relies on the following two technical lemmas.

\begin{lemma} \label{lemma:subterm_basic}
A proper subterm is a subterm that is not equal to the term. We have the following:
\begin{itemize}
    \item a) The subterm relation is transitive, i.e. if $t_1$ is a subterm of $t_2$ and $t_2$ is a subterm of $t_3$, then $t_1$ is a subterm of $t_3$.

    \item b) If $t_1$ is a proper subterm of $t_2$ and $t_2$ is a subterm of $t_3$, then $t_1$ is a proper subterm of $t_3$.

    \item c) The parsing of terms is unambiguous. More specifically if two terms $t_1$ and $t_2$ are identical, they have the same principal function symbol (say $f$). If we write $t_1=f(u_1,u_2,\ldots,u_d)$ and $t_2=f(u'_1,u'_2,\ldots,u'_d)$ then $u_j=u'_j$ for $j=1,2,\ldots,d$.
\end{itemize}
\end{lemma}

The proof of Lemma~\ref{lemma:subterm_basic} is easy and hence omitted.

\begin{lemma} \label{lemma:term_good}
Assume $t$ is a term and let $s_1,s_2,\ldots,s_\rho$ be a term-cut for $t$ such that $s_i$ is not a subterm of $s_j$ for all $i \neq j$. Let $u$ be a subterm of $t$, then there are two exclusive possibilities:

\begin{itemize}
    \item i) $s_1,s_2,\ldots,s_\rho$ is a term-cut for $u$.

    \item ii) $u$ is a proper subterm of some $s_j, j=1,2,\ldots,\rho$ ($u$ might be a subterm of more than one $s_j$).
\end{itemize}
\end{lemma}

\begin{IEEEproof}
We first show that the possibilities  i) and ii) are exclusive. Suppose on the contrary that a subterm $u$ satisfies both i) and ii). Then some $s_i$ is a proper subterm of $u$, and by Lemma~\ref{lemma:subterm_basic} it is a subterm of some $s_j$, which contradicts our assumption.

We now show that at least one of i) or ii) is satisfied. Assume $u$ is chosen as a subterm of $t$ that fails to satisfy both i) and ii) and such that it is not the proper subterm of some other subterm of $t$ which also fails i) and ii). Since the term $t$ satisfies i), $u$ must be a proper subterm of $t$. Thus $u$ occurs as a direct subterm in some subterm $v$ of $t$, i.e. $v$ can be written as $v=g(\ldots,u,\ldots)$ where $g$ is the principal function symbol in $v$. Since the subterm relation is transitive (Lemma~\ref{lemma:subterm_basic} part a), $v$ does not satisfy ii), and as $v=g(\ldots,u,\ldots)$ with $u$ not satisfying i), it follows from Lemma~\ref{lemma:subterm_basic} part c, that $v$ also fails to satisfy i). But this contradicts the assumption that $u$ was not a subterm of a subterm that failed to satisfy both i) and ii).
\end{IEEEproof}

We now prove Proposition~\ref{prop:two_cuts}.

\begin{IEEEproof}
Assume first that $C=\{s_1,s_2,\ldots,s_\rho\} \subseteq \Gamma_{\rm sub}$ is a term-cut for $\Gamma$. If some term $s_i$ is a subterm of $s_j$, remove $s_j$ from $C$. What remains after having repeated this procedure is a set $C' \subseteq C$ of minimal subterms in $C$. Clearly $C'$ is a term-cut for $\Gamma$. To keep the notation simple let us assume $C'=\{s_1,s_2,\ldots,s_{\rho'}\}$. It suffices to show that $C' \subseteq \Gamma_{\rm sub}$ is a cut that separates $S=\Gamma_{{\rm var}}$ from
$T=\Gamma$ in the directed graph $G_{\Gamma}$.

Assume there is a path $P$ from a variable $x \in \Gamma_{{\rm var}}$ to a term $t \in \Gamma$ which does not intersect $C'$. Since $C'$ satisfies the conditions for Lemma~\ref{lemma:term_good}, each term in $P$ satisfies either i) or ii). Since $x$ satisfies ii) while $t$ satisfies i), consider the first subterm $u$ in $P$ that satisfies i). The proper subterm of $u$ in $P$ satisfies ii), and hence is a proper subterm of some $s_j$. Thus $u$ must be identical to that $s_j$, since otherwise we could not express $u$ as a function of the subterms in the cut $C'$. This is not possible since $P$ was assumed not to intersect $C'=\{s_1,s_2,\ldots,s_{\rho'}\}$.

To prove the converse, assume that $C=\{s_1,s_2,\ldots,s_\rho\} \subseteq \Gamma_{\rm sub}$ is a cut that separates $S=\Gamma_{{\rm var}}$ from $T=\Gamma$ in the directed graph $G_{\Gamma}$.
Each subterm $t$ that does not belong to the cut $C$ and is on the same side of the cut as $T$ has each of its arguments  either in the cut or on the same side of the cut as $T$.  A simple argument by induction shows that subterms that  have all their arguments either in the cut or on the same side of the cut as $T$ can be written on the form $g(s_1,s_2,\ldots,s_\rho)$. This shows that $C$ defines a term-cut for $\Gamma$.

The fact that the size of a minimal term-cut is identical to the size of a minimal cut follows trivially from the first part of the proposition.
\end{IEEEproof}

According to the directed graph version of Menger's theorem \cite{Men27}, there exists a family $P$ of vertex-disjoint directed paths from $S = \Gamma_{{\rm var}}$ to $T = \Gamma$ and a vertex cut $C$ which consists of exactly one vertex from each path in $P$. Moreover, given the term set $\Gamma$, it is computationally feasible to find the exact value of the min-cut. By use of Dinic's algorithm \cite{Din70} for finding max flows in networks with unit capacities--which terminates in $O(V\sqrt{E})$ time--a term-cut of minimal size for any term set $\Gamma$ can be returned in time $O(|\Gamma_{\rm sub}|^2)$.

We would like to emphasize that any term-cut $C \subseteq \Gamma_{\rm sub}$ for $\Gamma$ is always a vertex cut that separates $S$ from $T$ {\em when $G_{\Gamma}$ is considered as a directed graph}. This is due to the antisymmetric nature of the subterm relation. Example \ref{example:directed_g_gamma} illustrates this distinction.

\begin{example} \label{example:directed_g_gamma}
Consider the term set
\[
    \Gamma=\left\{ h\Big(g\big(f(z),y\big),x\Big), l\big(f(z)\big),l(z)\right\}.
\]
The graph $G_{\Gamma}$ has vertex set
$V = \left\{x,y,z, f(z), l\big(f(z)\big), g\big(f(z),y\big), h\Big(g\big(f(z),y\big),x\Big)\right\}$, source set $S=\{x,y,z\}$ and target set $T=\left\{ h\Big(g\big(f(z),y\big),x\Big), l\big(f(z)\big), l(z)\right\}$ and is displayed in Figure~\ref{fig:directedB} a.

\begin{figure}
\begin{center}
    \includegraphics[scale=0.45]{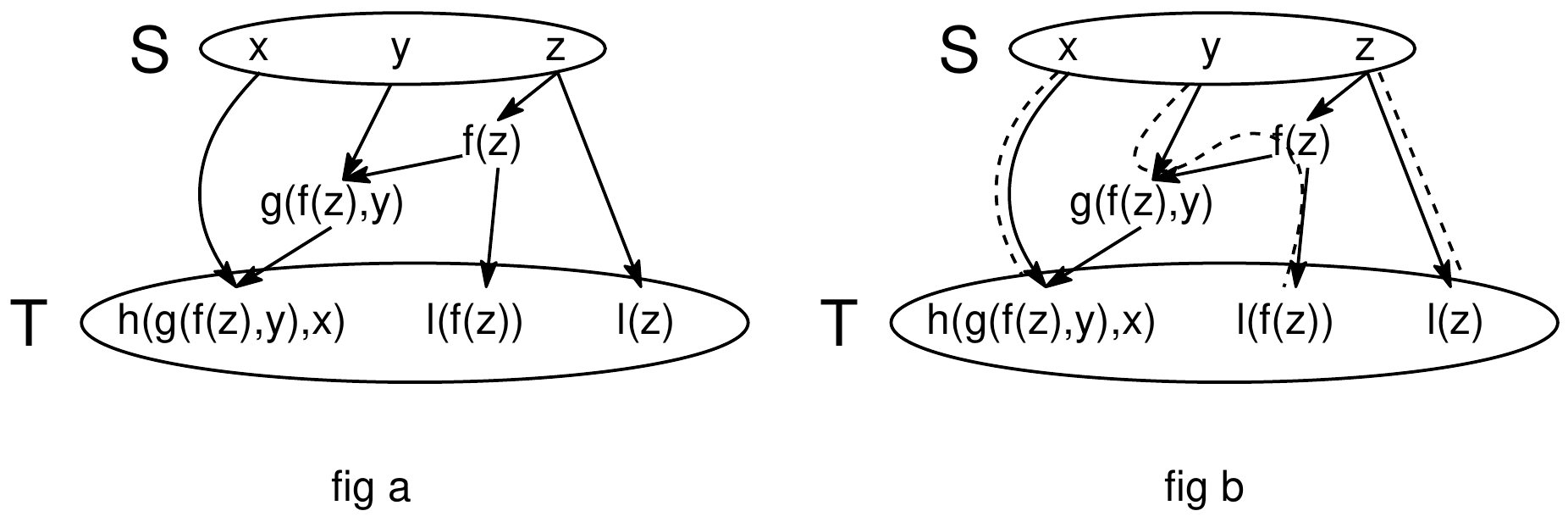}
\end{center}
\caption{Directed graph $G_\Gamma$.} \label{fig:directedB}
\end{figure}

Viewed as a directed graph, $G_{\Gamma}$ has a vertex cut (i.e. a term-cut for $\Gamma$) $C=\left\{ h\Big(g\big(f(z),y\big),x\Big), z\right\}$ of size $2$ and contains only $2$ vertex-disjoint directed paths from $S$ to $T$. Viewed as an undirected graph, $G_{\Gamma}$ has a minimal vertex cut of size $3$ and contains 3 vertex-disjoint undirected paths from $S$ to $T$ (see Figure~\ref{fig:directedB} b).
\end{example}

\subsection{Coding functions and their ability to disperse information in a single communication channel}

So far, we have treated function symbols as abstract entities; we now assign them explicit values.

\begin{definition}[Interpretation] \label{def:model}

Let $A$ be a finite set with $|A| \geq 2$, referred to as the alphabet. An {\em interpretation} for $\Gamma$ over $A$ is an assignment of the function symbols $\psi = \{\bar{f}_1,\bar{f}_2,\ldots,\bar{f}_l\}$, where $\bar{f}_i : A^{d_i} \rightarrow A$ for all $1 \leq i \leq l$.
\end{definition}

Once all the function symbols $f_i$ are assigned coding functions $\bar{f}_i$, then by composition each term $t_j \in \Gamma$ is assigned a function $\bar{t}_j : A^k \rightarrow A$. In order to simplify notations, we shall write functions by the way they map a tuple ${\bf a} = (a_1,a_2,\ldots,a_k) \in A^k$, and we typically write tuples in bold face. We shall abuse notations and also denote the {\em induced mapping} of the interpretation as $\psi : A^k \rightarrow A^r$, defined as
\begin{equation} \label{eq:psi} \nonumber
    \psi({\bf a}) = \big(\bar{t}_1({\bf a}), \bar{t}_2({\bf a}),\ldots,\bar{t}_r({\bf a})\big).
\end{equation}
Note that the definition of the induced mapping depends on the ordering of terms in $\Gamma$. However, our performance measures for interpretations and induced mappings will not depend on a particular ordering. 

\begin{example} \label{example:psi}
Consider $\Gamma_1$ introduced in Example~\ref{example:term-cut} and let $A = \mathbb{F}_2$. The interpretation $\psi = \{\bar{f}, \bar{g}, \bar{h}, \bar{m}\}$ given by
$\bar{f}(a_1,a_2) = a_1$, $\bar{g}(a_1,a_2) = a_1+a_2$, $\bar{h}(a_1,a_2,a_3) = a_2a_3+1$, $\bar{m}(a_1,a_2) = a_1a_2$ induces the mapping
\begin{equation} \nonumber
    \psi(a_1,a_2,a_3,a_4) = \big( (a_3+a_4)a_2+1, (a_3+a_4)a_2, a_1+a_3+a_4, a_3+a_4 \big).
\end{equation}
\end{example}

We are especially interested in how $\psi$ disperses its outputs, and how much information about the inputs can be obtained from the outputs. For any ${\bf b} \in A^r$, we denote the pre-image of ${\bf b}$ as ${\rm pre}({\bf b}) = \{ {\bf a} \in A^k : \psi({\bf a}) = {\bf b} \}$. The image and the one-to-one image of $\psi$ are respectively defined as
\begin{eqnarray*}
    {\rm image}(\psi) &:=& \left\{{\bf b} \in A^r: |{\rm pre}({\bf b})| \geq 1 \right\},\\
    {\rm one}(\psi) &:=& \left\{{\bf b} \in A^r: |{\rm pre}({\bf b})| = 1 \right\}.
\end{eqnarray*}
We now define the analogue of the value of a flow for information transfer on networks based on logic, which we refer to as the dispersion.

\begin{definition} \label{def:dispersion}
The {\em $\Gamma$-dispersion} and {\em one-to-one $\Gamma$-dispersion} of an interpretation $\psi$ for $\Gamma$ over $A$ are respectively defined as
\begin{eqnarray*}
    \gamma(\psi) &:=& \log_{|A|} |{\rm image}(\psi)|,\\
    \gamma_{{\rm one}}(\psi) &:=& \log_{|A|} |{\rm one}(\psi)|.
\end{eqnarray*}

We define the worst-case (average-case)  {\em $\Gamma$-dispersion} ({\em one-to-one $\Gamma$-dispersion}) of an interpretation $\psi$ for $\Gamma$ over $A$ with respect to the variables in $U \subseteq \Gamma_{{\rm var}}$ as the minimal (average) $\Gamma$-dispersion  (one-to-one $\Gamma$-dispersion) for any setting of the variables in $\Gamma_{{\rm var}} \setminus U$.
\end{definition}

We remark that since ${\rm one}(\psi) \subseteq {\rm image}(\psi)$, we have $\gamma_{{\rm one}}(\psi) \leq \gamma(\psi)$ for all interpretations $\psi$. For instance, the interpretations in Example~\ref{example:psi} has $\Gamma_1$-dispersion of $\log_2 6$, while $\psi({\bf a}+(0,0,1,1)) = \psi({\bf a})$ for all ${\bf a} \in A^4$ implies it has one-to-one $\Gamma_1$-dispersion $-\infty$.  

We finally define the (one-to-one) dispersion of $\Gamma$ over $A$ as the maximal $\Gamma$-dispersion (one-to-one $\Gamma$-dispersion, respectively) over all interpretations for $\Gamma$ over $A$, and we denote this value by $\gamma(\Gamma,|A|)$ (by $\gamma_{{\rm one}}(\Gamma,|A|)$, respectively) as this quantity clearly depends on $A$ via its cardinality only.  We say an interpretation $\psi$ has perfect dispersion if it for some finite alphabet equals the $\Gamma$-dispersion.

\begin{observation} \label{obs:no difference} 
For perfect dispersion there is no difference between worst case dispersion and average dispersion. Both are identical to the value of the minimal term cut with regards to the required variables.
\end{observation}

\section{Max-flow min-cut theorem for the dispersion of term sets} \label{sec:max-flow_min-cut}

The main purpose of this section is to prove the following max-flow min-cut theorem for the dispersion and the one-to-one dispersion of term sets.

\begin{theorem} [Max-flow min-cut theorem for dispersion] \label{th:dispersion}
Let $\Gamma$ be a term set with min-cut of $\rho$, then for any alphabet $A$,
\[
    \gamma_{{\rm one}}(\Gamma,|A|) \leq \gamma(\Gamma,|A|) \leq \rho.
\]

Conversely,
$$
    \lim_{|A| \rightarrow \infty} \gamma_{{\rm one}}(\Gamma,|A|) = \lim_{|A| \rightarrow \infty} \gamma(\Gamma,|A|) = \rho.
$$
\end{theorem}

The first part of the max-flow min-cut theorem is easily proved.

\begin{lemma} \label{lemma:max<min}
Let $\Gamma$ be a term set built on $k$ variables and with min-cut of $\rho \leq k$. Then for all $A$, $\gamma_{{\rm one}}(\Gamma,|A|) \leq \gamma(\Gamma,|A|) \leq \rho$. Furthermore, if $\rho < k$, then $\gamma_{{\rm one}}(\Gamma,|A|) \leq \log_{|A|}(|A|^\rho-1) < \rho$.
\end{lemma}

\begin{IEEEproof}
Let $C$ be a minimal term-cut for $\Gamma$. $C$ can be viewed as a term set, hence let $\psi_C$ be an interpretation for $C$ over $A$. The size of the image of its induced mapping is at most $|A|^\rho$. Furthermore, let $\psi_\Gamma$ be an interpretation for $\Gamma$ over $A$. Since all terms of $\Gamma$ can be expressed as functions of elements of $C$, the size of the image of $\psi_\Gamma$ is at most that of $\psi_C$, hence $|{\rm image}(\psi_\Gamma)| \leq |A|^\rho$ and $\gamma(\Gamma,|A|) \leq \rho$.

Furthermore, if $\rho < k$, the average number of pre-images per element of ${\rm image}(\psi_\Gamma)$ is at least $|A|^{k-\rho} > 1$. Therefore, there exists an element with more than one pre-image, and $|{\rm one}(\psi_\Gamma)| \leq |A|^\rho-1$.
\end{IEEEproof}

\subsection{Diversified term sets} \label{sec:diversifying}

We first prove the max-flow min-cut theorem for the dispersion in the specific case where each subterm has a distinct function symbol. More specifically, we define the diversified term set by assigning a new function symbol to each subterm that is not a variable.

\begin{definition}[Diversified term set] \label{def:diversified}
For any term set $\Gamma$, the {\em diversified term set} $\Gamma^{\rm div}$ is built on the same variables as $\Gamma$ and its function symbols are obtained by replacing the principal function $g$ of any $u \in \Gamma_{\rm sub} \backslash \Gamma_{\rm var}$ by a new function symbol $g_u$ of the same arity as $g$.
\end{definition}

\begin{example} \label{example:diversified}
Recall the term set $\Gamma_1$ from Example~\ref{example:term-cut}:
\[
    \Gamma_1= \left\{h\big(f(x,y), g(z,w), f(y,x)\big), m\big(g(z,w), f(y,x)\big), g\big(f(x,y), g(z,w)\big),  f\big(g(z,w), f(y,x)\big) \right\},
\]
then
\begin{eqnarray*}
    \Gamma_1^{{\rm div}} &=& \left\{h_{h(f(x,y),g(z,w),f(y,x))}\big(f_{f(x,y)}(x,y), g_{g(z,w)}(z,w), f_{f(y,x)}(y,x)\big),\right.\\
    && m_{m(g(z,w),f(y,x))}\big(g_{g(z,w)}(z,w), f_{f(y,x)}(y,x)\big),\\
    && \left. g_{g(f(x,y),g(z,w))}\big(f_{f(x,y)}(x,y),g_{g(z,w)}(z,w)\big),
    f_{f(g(z,w),f(y,x))}\big(g_{g(z,w)}(z,w),f_{f(y,x)}(y,x)\big)\right\}.
\end{eqnarray*}
We can simplify the indices and rewrite the diversified term set as
\[
    \left\{ h \big(f_1(x,y),g_1(z,w),f_2(y,x) \big),
    m\big(g_1(z,w),f_2(y,x)\big),
    g_2\big(f_1(x,y),g_1(z,w)\big),
    f_3\big(g_1(z,w),f_2(y,x)\big) \right\}.
\]
\end{example}

We remark that before diversification, the same function symbol may be assigned to different subterms (e.g., $f(x,y)$ and $f(y,x)$ have the same principal function symbol in $\Gamma_1$). However, after diversification, there cannot be such overlap, as each subterm is assigned a distinct principal function. By definition, it is easily seen that the graph $G_{\Gamma^{{\rm div}}}$ is isomorphic to $G_{\Gamma}$. In particular, $\Gamma$ and $\Gamma^{{\rm div}}$ have the same min-cut.

For diversified term sets, maximal dispersion can be achieved via {\em routing}, which is defined in a similar way to the case of ordinary networks. Let $\Gamma = \{t_1,t_2,\ldots,t_r\}$ be built on the variables $\{x_1,x_2,\ldots,x_k\}$ and have min-cut of $\rho$. Let $P$ be a set of $\rho$ vertex-disjoint paths from $\Gamma_{{\rm var}}$ to $\Gamma$ in $G_\Gamma$ which, without loss, start in $x_1,x_2,\ldots,x_\rho$ and end in $t_1,t_2,\ldots,t_\rho$, respectively.

\begin{definition}[Routing] \label{def:routing}
A distinct function symbol $g_v$ is associated to each subterm $v \in \Gamma_{\rm sub}$. 
If $u_j$ is the direct subterm of $v$ on the same path, then we let $\bar{g}_v(a_1,a_2,\ldots,a_d) = a_j$. Otherwise, i.e. if $v$ does not belong to any path in $P$, then $\bar{g}_v(a_1,a_2,\ldots,a_d) =1$.
\end{definition}

Note that our definition of routing depends on the set of paths $P$, and hence is not unique. However, the dispersion and one-to-one dispersion of routing do not depend on the choice of $P$. It is straightforward to verify that using routing, all points of the form $(a_1,a_2,\ldots,a_\rho,1,\ldots,1) \in A^k$ are mapped to $(a_1, a_2, \ldots, a_\rho,1, \ldots, 1) \in A^r$, thus yielding a $\Gamma$-dispersion of $\rho$. Furthermore, when $\rho = k$, the induced mapping (restricted to the first $\rho$ coordinates) becomes the identity on $A^\rho$ and hence $\gamma_{{\rm one}}(\Gamma,|A|)=\rho$. However, routing has one-to-one dispersion $- \infty$ when $\rho < k$. In order to thwart this drawback, we define {\em one-to-one routing} below.

\begin{definition}[One-to-one routing] \label{def:1-1_routing}
Let $v$ be a subterm of the form $v = g_v(u_1,u_2,\ldots,u_d)$, and denote the set of arguments equal to variables $x_{\rho+1},x_{\rho+2},\ldots,x_k$ as $u_{i_1}, u_{i_2}, \ldots, u_{i_m}$. We define the coding function $\bar{g}_v:A^d \rightarrow A$ as follows. If a path in $P$ goes through $v$, denote the direct subterm of $v$ on the same path as $u_j$; then, if $a_{i_1} = a_{i_2} = \ldots = a_{i_m} = 1$, we let $\bar{g}_v(a_1,a_2,\ldots,a_d)=a_j$. Otherwise, let $\bar{g}_v(a_1,a_2,\ldots,a_d) = 1$.
\end{definition}

With one-to-one routing, it is straightforward to check that the $(|A|-1)^\rho$ points of the form
\\
$(a_1,a_2,\ldots,a_{\rho},1,\ldots,1) \in A^k$ with $a_{1} \neq 1, a_{2} \neq 1, \ldots,a_{\rho} \neq 1$ are mapped in a one-to-one fashion to $(a_1, a_2, \ldots, a_\rho,1, \ldots, 1) \in A^r$, thus yielding a one-to-one $\Gamma$-dispersion of at least $\rho \log_{|A|}(|A|-1)$.

We obtain the following max-flow min-cut result for diversified term sets.

\begin{proposition} \label{prop:dispersion_diversified}
Assume $\Gamma$ is a term set built on $k$ variables and with min-cut $\rho$. Let $A$ be an alphabet of size $|A| \geq 2$, then $\gamma(\Gamma^{{\rm div}},|A|)=\rho$, and it is achieved by routing. Furthermore, if $\rho = k$, $\gamma_{{\rm one}}(\Gamma^{{\rm div}},|A|)=\rho$ is achieved by routing, while if $\rho<k$, one-to-one routing yields
$$
    \rho \log_{|A|}(|A|-1) \leq \gamma_{{\rm one}}(\Gamma^{{\rm div}},|A|) \leq \log_{|A|}(|A|^\rho-1).
$$
\end{proposition}

\subsection{Dynamic routing} \label{sec:dynamic_headers}

The construction of coding functions in Section~\ref{sec:diversifying} used the fact that each subterm $v$ was assigned a distinct function symbol. However, in general distinct subterms might be assigned the same function symbol (e.g., $f(x,y)$ and $f(y,x)$). The proof of the general case relies on {\em dynamic routing}, defined below.

For $|A| > |\Gamma_{\rm sub}|$, there exist two sets $B$ and $R$ with $1 \leq |R|  \leq |\Gamma_{\rm sub}|$ such that $|A|=|(\Gamma_{\rm sub} \times B) \cup R|$ where the union is disjoint. We shall abuse notation slightly and assume $A = (\Gamma_{\rm sub} \times B) \cup R$. By construction, a tuple ${\bf a} = (a_1,a_2,\ldots,a_{d_j}) \in A^{d_j}$ either has an element in $R$ or has each $a_i=(u_i, b_i) \in \Gamma_{\rm sub} \times B$.

\begin{definition}[Dynamic routing] \label{def:dynamic}
Consider the term set $\Gamma^{{\rm div}}$ first, which contains one function symbol $g_v$ for each subterm $v \in \Gamma_{\rm sub}$. Select coding functions $\bar{g}_v$ over $B$ using routing, as in Definition~\ref{def:routing}. We then define the functions $\bar{f}_j(a_1,a_2,\ldots,a_{d_j})$ over $A$ as follows. If each $a_i$ is of the form $a_i=(u_i, b_i) \in \Gamma_{\rm sub} \times B$, let $s$ denote the term $s=f_j(u_1, u_2,\ldots,u_{d_j})$; then if $s \in \Gamma_{\rm sub}$
\[
    \bar{f}_j(a_1,a_2,\ldots,a_{d_j})=(s, \bar{g}_{s}(b_1,b_2,\ldots,b_{d_j})) \in \Gamma_{\rm sub} \times B.
\]
Otherwise, let $\bar{f}_{j}(a_1,a_2,\ldots,a_{d_j})=r$ for some $r \in R$.
\end{definition}

We can similarly define {\em dynamic one-to-one routing}. Remark that the headers $u_i$ of the inputs then indicate to the coding function $\bar{f}_j$ which subterm $v$ it is located on, and hence which function $\bar{g}_v$ to use. We say an input message $a_j$ for the variable $x_j$ is {\em correctly formatted} if it is of the form $(x_j,b_j)$ where $b_j \in B$, and we denote the set of all correctly formatted inputs as
$I := \{{\bf a} \in (\Gamma_{{\rm var}} \times B)^k : a_j = (x_j,b_j)\}.$
Moreover, the set of correctly formatted outputs is denoted as
$
    J := \{{\bf a} \in (\Gamma \times B)^r : a_j = (t_j,b_j)\},
$
and for all ${\bf a} \in J$, we denote the data part of ${\bf a}$ as $b({\bf a}) = (b_1,b_2,\ldots,b_r) \in B^r$. The idea behind dynamic routing is that if all inputs are correctly formatted (i.e. have the correct headers) then the coding functions $\bar{f}_j$ mimic the behavior of the routing functions $\bar{g}_v$. Thus, correctly formatted messages in $I$ are mapped--in a one-to-one fashion if one-to-one routing is used--to correctly formatted outputs in $J$ (as long as they are mapped by the functions $\bar{g}_v$), while other messages will be mapped to an ``error message'' in $R$. We obtain the following lemma.

\begin{lemma} \label{lemma:dynamic}
Let $\psi$ be a dynamic routing interpretation for $\Gamma$ over $A$ based on the routing interpretation $\phi$ for $\Gamma^{{\rm div}}$ over $B$, then $\{{\bf a} \in J: b({\bf a}) \in {\rm image}(\phi)\} \subseteq {\rm image}(\psi)$. Similarly, if $\psi_{\rm one}$ is a dynamic one-to-one routing interpretation for $\Gamma$ over $A$ based on the one-to-one routing interpretation $\phi_{\rm one}$ for $\Gamma^{{\rm div}}$ over $B$, then $\{{\bf a} \in J: b({\bf a}) \in {\rm one}(\phi_{\rm one})\} \subseteq {\rm one}(\psi_{\rm one})$.
\end{lemma}

Lemma~\ref{lemma:dynamic}, together with Proposition~\ref{prop:dispersion_diversified}, gives a lower bound on the dispersion and one-to-one dispersion. By choosing an appropriate alphabet size, we can prove the following quantitative version of the max-flow min-cut theorem for the dispersion.

\begin{theorem} \label{th:dispersion_refined}
Let $\Gamma$ be a term set built on $k$ variables and with min-cut of $\rho$. For $\epsilon < \rho$, let
$n_1 := |\Gamma_{\rm sub}|^{\rho/\epsilon} (1 - |\Gamma_{\rm sub}|^{1 - \rho/\epsilon})^{-\rho/\epsilon}$.
Then for all $|A| \geq n_1$, $\gamma(\Gamma,|A|) \geq \rho - \epsilon$ and if $\rho = k$, $\gamma_{{\rm one}}(\Gamma,|A|) \geq \rho - \epsilon$. These are achieved by dynamic routing.

Moreover, for $\epsilon < \frac{\rho}{1 + \log_{|\Gamma_{\rm sub}|} 2}$, let
$n_2 := |\Gamma_{\rm sub}|^{\rho/\epsilon} (1 - 2|\Gamma_{\rm sub}|^{1 - \rho/\epsilon})^{-\rho/\epsilon}$.
then for all $|A| \geq n_2$, if $\rho < k$, $\gamma_{{\rm one}}(\Gamma,|A|) \geq \rho - \epsilon$ is achieved by dynamic one-to-one routing.
\end{theorem}

\begin{IEEEproof}
We only prove the case involving $n_2$, the other being proved similarly. Suppose $A$ is an alphabet with $|A| \geq n_2$ and let $\psi$ be the mapping induced by dynamic one-to-one routing for $\Gamma$ over $A$. By Lemma~\ref{lemma:dynamic} and Proposition~\ref{prop:dispersion_diversified}, $|{\rm one}(\psi)| \geq (|B|-1)^\rho$.
We have
\begin{equation} \nonumber
    \frac{|B|-1}{|A|^{1-\epsilon/\rho}} \geq \frac{|A|^{\epsilon/\rho}}{|\Gamma_{\rm sub}|} \left(1 - \frac{2|\Gamma_{\rm sub}|}{|A|}\right)
    \geq \frac{|A|^{\epsilon/\rho}}{|\Gamma_{\rm sub}|} \left(1 - \frac{2|\Gamma_{\rm sub}|}{|\Gamma_{\rm sub}|^{\rho/\epsilon}}\right)
    \geq 1,
\end{equation}
where the successive inequalities follow from $|B| \geq \frac{|A|}{|\Gamma_{\rm sub}|} - 1$, $n_2 \geq |\Gamma_{\rm sub}|^{\rho/\epsilon}$, and the definition of $n_2$, respectively. Thus $\gamma_{{\rm one}}(\Gamma, |A|) \geq \log_{|A|}(|B|-1)^\rho \geq \rho - \epsilon$.
\end{IEEEproof}

\section{Max-flow min-cut for the R\'enyi entropy} \label{sec:renyi}

\subsection{R\'enyi entropy of an interpretation} \label{sec:entropy_model}

Let $\Gamma$ be a set of $r$ terms built on $k$ variables, and let $\psi$ be an interpretation for $\Gamma$ over an alphabet $A$. Once $A$ and $\psi$ are fixed, the flow of data from the inputs ${\bf a} \in A^k$ to the outputs $\psi({\bf a}) \in A^r$ can be viewed as the transmission of a random variable $\bar{\bf a}$ taking values in $A^k$ through the deterministic channel operating the induced mapping $\psi$. Its capacity $C_\psi$ is easily computed: denoting the mutual information between two random variables $X$ and $Y$ as $I(X;Y)$, we have
\begin{eqnarray} \nonumber
    C_\psi &:=& \sup_{\bar{\bf a}} I(\bar{\bf a}; \psi(\bar{\bf a}))\\
    \label{eq:capacity}
     &=& \sup_{\bar{\bf a}} H(\psi(\bar{\bf a})) = \gamma(\psi),
\end{eqnarray}
where $H$ denotes the Shannon entropy. The maximum is reached when $\bar{\bf a}$ has the following probability distribution: for each ${\bf b} \in {\rm image}(\psi)$, select $a({\bf b}) \in {\rm pre}({\bf b})$ and let $\mathbb{P}\{\bar{\bf a} = a({\bf b})\} = |{\rm image}(\psi)|^{-1}$. Eq. (\ref{eq:capacity}) shows that the capacity of the channel is given by the dispersion of the interpretation considered. This justifies our study of the dispersion in Section \ref{sec:max-flow_min-cut}. Thus, the max-flow min-cut theorem for the dispersion states that the channel capacity asymptotically converges to the min-cut $\rho$ of the term set, i.e.
\begin{equation} \nonumber
    \sup_{A,\psi} C_\psi = \sup_{A,\psi} \gamma(\psi) = \rho.
\end{equation}

We note that the capacity $C_\psi$ is achieved for a specific input random variable which is not uniformly distributed over all inputs. This represents the capacity achieved when the sources are {\em cooperative} and agree on a coding scheme for the input. We are now interested in the case where the sources are non-cooperative, and as such we assume that the inputs are uniformly distributed over $A^k$.

This opens the question of the most accurate measure of performance for a term set. If the input $\bar{\bf a} \in A^k$ is uniformly distributed, then $\psi(\bar{\bf a})$ is a random variable with values in $A^r$, where for all ${\bf b} \in A^r$
\begin{equation} \nonumber
    p_{\bf b} = \mathbb{P}\{ \psi(\bar{\bf a}) = {\bf b} \} = \frac{|{\rm pre}({\bf b})|}{|A|^k}.
\end{equation}
The normalized R\'enyi entropy over $A$ of order $0 \leq \alpha \leq \infty$ of the random variable $\psi(\bar{\bf a})$, which we will simply denote as $H_\alpha(\psi)$, is thus given by \cite{Ren61}
\begin{eqnarray*}
    H_{\alpha}(\psi)&:=&\frac{1}{1-\alpha} \log_{|A|}\sum_{{\bf b} \in A^r} p^{\alpha}_{\bf b}\\
    &=& \frac{\alpha}{\alpha-1} \left(k - \frac{1}{\alpha} \log_{|A|} \sum_{{\bf b} \in A^r}|{\rm pre}({\bf b})|^\alpha \right) \quad 0 < \alpha < 1 \, {\rm or} \, 1 < \alpha < \infty.
\end{eqnarray*}

Three further special cases need close attention.

First, when $\alpha=0$, the R\'enyi entropy is the logarithm of the cardinality of the number of outcomes, which is often referred to as the {\em Hartley entropy}. In our case, the Hartley entropy of $\psi$ is its dispersion, which is equal to the channel capacity:
$$
    H_0(\psi) := \log_{|A|} |\{{\bf b} \in A^r: p_{\bf b} > 0\}|
    = \gamma(\psi).
$$

Second, we remark that $\log_{|A|} |{\rm pre}({\bf b})|$ is the uncertainty about the input when the message ${\bf b}$ is received. Hence the variable $k - \log_{|A|} |{\rm pre}({\bf b})|$ is the amount of information (counted in symbols in $A$) that can be inferred about ${\bf a}$ from ${\bf b} = \psi({\bf a})$. The {\em Shannon entropy}, obtained when $\alpha=1$, is therefore the expected amount of information inferred from the term set about the input messages:

$$
    H_1(\psi) := -  \sum_{{\bf b} \in A^r} p_{\bf b} \log_{|A|} p_{\bf b}
    = \mathbb{E}\left\{k - \log_{|A|} |{\rm pre}({\bf b})|\right\}.
$$

Third, when $\alpha=\infty$, the {\em min-entropy} quantifies the amount of information that can be inferred from any output, by considering the point ${\bf b} \in A^r$ with the most pre-images:

$$
    H_\infty(\psi) := - \log_{|A|} \max_{{\bf b} \in A^r} p_{\bf b}
    = \min_{{\bf b} \in A^r}\left\{k - \log_{|A|} |{\rm pre}({\bf b})|\right\}.
$$

Note that there exist interpretations $\psi_1$ and $\psi_2$ such that $H_0(\psi_1) > H_0(\psi_2)$ and yet $H_\alpha(\psi_1) < H_\alpha(\psi_2)$ for some $\alpha>0$. Therefore, having the highest dispersion does not guarantee to perform well for the other measures.

\subsection{R\'enyi entropy and one-to-one dispersion} \label{sec:renyi_one}

Although the dispersion is a special R\'enyi entropy, Proposition \ref{prop:one-to-one_v_renyi} below shows that the one-to-one dispersion cannot be viewed as a R\'enyi entropy. Even more strikingly, the second statement shows that the one-to-one dispersion can actually conflict with the other entropy measures. This can be intuitively explained as follows. The R\'enyi entropies measure to which degree the inputs have been mixed by the induced mapping $\psi$, for instance a high min-entropy guarantees that not too many inputs have been mapped to the same output. On the other hand, in order to guarantee a high one-to-one dispersion, the mixing has to be controlled so that one output is reserved for each input in the one-to-one pre-image. This control may significantly reduce the entropy if the size of the one-to-one image is very close to the total number of images, which occurs necessarily if the former is very near $|A|^\rho$.

\begin{proposition} \label{prop:one-to-one_v_renyi}
First, for any term set $\Gamma$, there exists an interpretation $\psi$ for $\Gamma$ over any alphabet $A$ such that $\gamma_{\rm one}(\psi) < H_\infty(\psi)$. Second, let $\Gamma_1 = \{f(x,y)\}$ be a term set with min-cut $1$. Then for any $\alpha > 0$, there exists an interpretation $\psi$ for $\Gamma_1$ such that $\gamma_{\rm one}(\psi) > H_\alpha(\psi)$. Furthermore, if an interpretation $\psi$ for $\Gamma_1$ has maximal one-to-one $\Gamma_1$-dispersion, then $H_1(\psi)$ tends to zero for large $|A|$. Conversely, if $H_\alpha(\psi) = 1$ for some $\alpha > 0$ then $\gamma_{\rm one}(\psi) = - \infty$.
\end{proposition}

\begin{IEEEproof}
First, if all the coding functions are constant, $\gamma_{\rm one}(\psi) = -\infty$. Second, one-to-one routing has maximal one-to-one $\Gamma_1$-dispersion equal to $\log_{|A|}(|A|-1)$ which tends to $1$, while its R\'enyi entropy is given by
$$
    \frac{1}{1-\alpha} \log_{|A|} \{|A|-1 + (|A|^2-|A|+1)^\alpha\} - \frac{2\alpha}{1-\alpha},
$$
which tends to $1 - \frac{\alpha}{1-\alpha}$ for $\alpha < \frac{1}{2}$. Conversely, if $H_\alpha(\psi) = 1$ then all outputs have exactly $|A|$ pre-images.
\end{IEEEproof}

\subsection{Max-flow min-cut theorem for the R\'enyi entropy}

Similarly to the dispersion, we denote the maximum R\'enyi entropy over all interpretations for $\Gamma$ over $A$ as $H_\alpha(\Gamma, |A|)$. The max-flow min-cut theorem for the dispersion indicates that the R\'enyi entropy for $\alpha = 0$ tends to the min-cut of the term set. We shall prove the following result.

\begin{theorem}[Max-flow min-cut theorem for the R\'enyi entropy] \label{th:renyi}
Let $\Gamma$ be a term set with min-cut of $\rho$, then $$\lim_{|A| \rightarrow \infty} H_\alpha(\Gamma,|A|) = \rho \quad \mbox{for all}\, 0 \leq \alpha < 1.$$ Conversely, for any $\alpha > 1$, there exists a term set $\Gamma$ with min-cut $\rho$ for which $\lim_{|A| \rightarrow \infty} H_\alpha(\Gamma,|A|) < \rho.$
\end{theorem}

The max-flow min-cut theorem for the R\'enyi entropy with $\alpha < 1$ is actually based on dynamic routing.

\begin{proposition}[Max-flow min-cut theorem for $\alpha < 1$]\label{prop:alpha<1}
Let $\Gamma$ be a term set built on $k$ variables and with min-cut of $\rho$ and for all $0 < \alpha < 1$ define $\beta = \rho + \frac{\alpha}{1-\alpha}k$ and $n_3 = (2|\Gamma_{\rm sub}|)^{\frac{\epsilon}{\beta}}.$ Then for any alphabet $A$ with $|A| \geq n_3$, $H_\alpha(\Gamma,|A|) \geq \rho - \epsilon$, which is achieved by dynamic routing.
\end{proposition}

\begin{IEEEproof}
Suppose that dynamic routing is used over an alphabet $A$ (recall the notations from Section \ref{sec:dynamic_headers}). It is clear that any output of the form $((t_1,b_1),(t_2,b_2),\ldots,(t_\rho,b_\rho),(t_{\rho+1},1),\ldots,(t_r,1)) \in (\Gamma \times B)^r$ has exactly $|B|^{k-\rho}$ pre-images, namely those of the form $((x_1,b_1), (x_2,b_2), \ldots, (x_k, b_k)) \in (\Gamma_{\rm var} \times B)^k$, where $b_{\rho+1}, b_{\rho+2}, \ldots, b_k \in B$. Let us denote this set of $|B|^\rho$ outputs as $C$ and compute the R\'enyi entropy of dynamic routing.
\begin{eqnarray*}
    H_\alpha(\psi) &=& \frac{1}{1-\alpha} \log_{|A|} \sum_{{\bf b} \in A^r} |{\rm pre}({\bf b})|^\alpha - k \frac{\alpha}{1-\alpha}\\
    &\geq& \frac{1}{1-\alpha} \log_{|A|} \sum_{{\bf b} \in C} |B|^{\alpha(k-\rho)} - k \frac{\alpha}{1-\alpha}\\
    &=& \rho - \beta \log_{|A|}\left\{|\Gamma_{\rm sub}| + \frac{|R|}{|B|}\right\}\\
    &\geq& \rho - \beta \log_{|A|}(2|\Gamma_{\rm sub}|)\\
    &\geq& \rho - \epsilon.
\end{eqnarray*}
\end{IEEEproof}

The R\'enyi entropy for large $\alpha$ is sensitive to some types of bottlenecks which cannot be handled with the graphic approach. Indeed, we design below a family of term sets for which the R\'enyi entropy does not tend to the min-cut for $\alpha > 1$ and arbitrarily close to $1$. Therefore, there is no max-flow min-cut theorem for the R\'enyi entropy with $\alpha > 1$.

For all $k \geq 2$, we define the set $\Gamma_k$ of $k^2$ terms, built on $k^2$ variables $x^a_j$, $0 \leq a,j \leq k-1$ and on one function symbol $f$ of arity $k$, to be
\begin{equation} \nonumber
    \Gamma_k = \left\{ t_{i,j} = f(x_i^0, x_j^1,x_j^2,\ldots,x_j^{k-1}) : 0 \leq i,j \leq k-1 \right\}.
\end{equation}

\begin{proposition} [No max-flow min-cut theorem for $\alpha > 1$] \label{prop:no_theorem}
The term set $\Gamma_k$ has min-cut of $k^2$. However, for $\alpha > \frac{k}{k-1}$ and all $A$,
$$
    H_\alpha(\Gamma_k,|A|) \leq \frac{(2k-1)\alpha-k}{\alpha-1} < k^2.
$$
\end{proposition}

\begin{IEEEproof}
First, we prove that $\Gamma_k$ has min-cut of $k^2$ by constructing $k^2$ vertex-disjoint paths from $\Gamma_{k,{\rm var}}$ to $\Gamma_k$. We have $x_j^0 \in t_{j,j}$ for all $j$ and if $a \geq 1$, $x_j^a \in t_{b,j}$ for any $0 \leq b \leq k-1$. Therefore, $(x_j^0,t_{j,j})$ for all $j$ and $(x_j^a,t_{a+j \mod k,j})$ for all $a \geq 1$ and $j$ form a set of $k^2$ vertex-disjoint paths.

Second, we give an upper bound on the R\'enyi entropy of any interpretation $\psi$ for $\Gamma_k$ over an alphabet $A$. Consider the set $C = \{{\bf a} \in A^{k^2} : a_0^0 = a_1^0 = \ldots = a_{k-1}^0\}$, then if ${\bf a} \in C$, $\bar{t}_{i,j}({\bf a}) = \bar{t}_{i',j}({\bf a})$ for all $i,i'$, and $j$ and hence only $k$ terms are non necessarily equal. Therefore, $|C| = |A|^{k^2-k+1}$ while the size of the image of $C$ is at most $|A|^k$. We have
\begin{eqnarray}
    \nonumber
    H_\alpha (\psi) &=& k^2 \frac{\alpha}{\alpha-1} - \frac{1}{\alpha-1} \log_{|A|} \sum_{{\bf b} \in A^{k^2}} |{\rm pre}({\bf b})|^\alpha\\
    \nonumber
    &\leq& k^2 \frac{\alpha}{\alpha-1} - \frac{1}{\alpha-1} \log_{|A|} \sum_{{\bf b} \in \psi(C)} |{\rm pre}({\bf b})|^\alpha\\
    \label{eq:alpha1}
    &\leq& k^2 \frac{\alpha}{\alpha-1} - \frac{1}{\alpha-1} \log_{|A|} \{ |A|^k |A|^{\alpha(k-1)^2} \}\\
    \nonumber
    &=& \frac{(2k-1)\alpha-k}{\alpha-1},
\end{eqnarray}
where (\ref{eq:alpha1}) follows the fact that since $\alpha > 1$, the summation is minimized when all terms are equal.
\end{IEEEproof}

The case of the Shannon entropy ($\alpha = 1$) remains open. This is an important question, as a max-flow min-cut theorem for the Shannon entropy would mean that the amount of information obtained in the non-cooperative case is asymptotically equal to that in the cooperative case. We would like to highlight the difficulty of treating the Shannon entropy case. It can be shown that for any fixed $n$ and for any $\epsilon > 0$, there exists $0<\alpha < 1$ such that $H_1(X) \geq H_\alpha(X) - \epsilon$ for any probability distribution $X$ on $n$ points. However, we show below that $\epsilon$ cannot be chosen independently of $n$.

\begin{example}
For $0<\alpha < 1$ and for  $n\geq 2$ consider the probability distribution $X=X_{\alpha, n}$ given by:
\[
    p_1=p_2=\ldots=p_{n-1}=\frac{1-\alpha}{n-1}, \quad p_n=\alpha.
\]
We obtain
\begin{eqnarray*}
    H_1(X) &=& -(1-\alpha) \log_n \left\{\frac{1-\alpha}{n-1}\right\} - \alpha \log_n \alpha, \\
    H_{\alpha}(X) &=& \frac{1}{(1-\alpha)} \log_n \left\{(n-1) \left(\frac{1-\alpha}{n-1} \right)^\alpha + \alpha^\alpha \right\}.
\end{eqnarray*}
It is not hard to show that $\lim_{n \rightarrow \infty} H_1(X) = 1-\alpha$ while $\lim_{n \rightarrow \infty} H_{\alpha}(X)= 1$, and hence
\[
    \lim_{\alpha \rightarrow 1} \lim_{n \rightarrow \infty} H_1(X) = 0 \neq 1 = \lim_{\alpha \rightarrow 1} \lim_{n \rightarrow \infty} H_\alpha(X).
\]
\end{example}

In contrast, we remark that the diversified term sets satisfy a much more general max-flow min-cut theorem. Indeed, when using routing, the number of pre-images of any output is a constant given by $|A|^{k - \rho}$. We obtain the following result.

\begin{proposition} \label{prop:renyi_diversified}
For any term set $\Gamma$ with min-cut of $\rho$, any $\alpha$, and any alphabet $A$,
$$
    H_\infty (\Gamma^{\rm div},|A|) = H_\alpha (\Gamma^{\rm div},|A|) = \rho.
$$
\end{proposition}

Proposition \ref{prop:renyi_diversified} shows that the case of traditional network coding, where distributed coding functions are absent, is trivial. This motivates our study of term sets with distributed coding functions, which yield different types of bottlenecks that cannot be captured by the typical directed graph approach.

\section{Dispersion of linear coding functions} \label{sec:linear}

\subsection{Insufficiency of linear coding functions} \label{sec:linear_v_non-linear}

We now consider the important class of linear coding functions. First, {\em scalar linear} functions are defined when $A$ is organized as a field $\mathbb{F}_q$ for some prime power $q$, and the coding functions $\bar{f} : \mathbb{F}_q^d \rightarrow \mathbb{F}_q$ can be written as $\bar{f}(a_1,a_2,\ldots,a_d)=\sum_{i=1}^da_ib_i$ for $b_1,b_2,\ldots,b_d \in \mathbb{F}_q$. A more general class are the {\em matrix linear} coding functions that are defined when $A$ is organized as a finite vector space $V$  and the coding functions $\bar{f} : V^d \rightarrow V$ can be written as $\bar{f}(a_1,a_2,\ldots,a_d)= \sum_{i=1}^d F_ia_i$ where $F_1,F_2,\ldots,F_d$ are linear maps from $V$ to $V$.

Clearly, if the coding functions are linear, then so is the induced mapping of the corresponding interpretation. The structure of linear maps allows us to characterize their one-to-one dispersion in Proposition~\ref{prop:all_nothing} below.

\begin{proposition} \label{prop:all_nothing}
For any set $\Gamma$ of terms built on $k$ variables and with min-cut $\rho$ and for any linear interpretation $\psi$ for $\Gamma$, then $\gamma_{{\rm one}}(\psi) = \rho$ if and only if $\gamma(\psi) = \rho = k$ and $\gamma_{{\rm one}}(\psi) = -\infty$ otherwise.
\end{proposition}

\begin{IEEEproof}
Let $A$ be a vector space of dimension $r$ over a finite field $F$, and let $\psi$ be a linear interpretation for $\Gamma$ over $A$. Since the induced mapping is linear, each point in its image has exactly $|\ker(\psi)|=|F|^d$ pre-images, where $d=\dim(\ker(\psi))$.

If $d=0$, then each point in the image of $\psi$ has exactly one pre-image, and hence $\gamma_{{\rm one}}(\psi)=k$. Since $\gamma_{{\rm one}}(\psi) \leq \rho$ by Lemma~\ref{lemma:max<min}, while $\rho \leq k$, we obtain $\gamma_{{\rm one}}(\psi)=k=\rho$. Conversely, it is easily shown that if $\gamma(\psi)=\rho=k$, then $\gamma_{{\rm one}}(\psi) = \rho$. If $d \geq 1$, then each point in the image of $\psi$ has more than one pre-image and $\gamma_{{\rm one}}(\psi)=-\infty$.
\end{IEEEproof}

Because linear maps disperse information uniformly, the R\'enyi entropy of a linear map does not depend on the coefficient $\alpha$ and is hence equal to its dispersion: for any linear interpretation $\psi$ for $\Gamma$ over $A$, we have $H_\infty(\psi) = H_\alpha(\psi) = \gamma(\psi)$ for all $0 \leq \alpha \leq \infty$. This shows a clear limitation of linear maps, as Theorem \ref{th:renyi} shows that the min-entropy may not reach the min-cut. In particular, for the term set $\Gamma_k$ introduced for Proposition \ref{prop:no_theorem}, the min-cut is equal to $k^2$ while the min-entropy of any interpretation (and hence the dispersion of any linear interpretation) is upper bounded by $2k-1$.

As Theorem~\ref{th:dispersion} indicates, although it is not always possible to reach the min-cut for any fixed finite alphabet, this can be achieved asymptotically. The class of term sets $\Gamma$ can then naturally be divided into two disjoint classes whether there exist coding functions that achieve perfect dispersion equal to the min-cut. If $\Gamma$ has perfect dispersion, we also say $\Gamma$ is {\em solvable} and a {\em solution} is an interpretation with dispersion equal to the min-cut. Solvable term sets are easily found; conversely, we implicitly proved that the term sets $\Gamma_k$ are not solvable for all $k$.

More generally, the dispersion of a scalar linear interpretation is always an integer, therefore if a term set $\Gamma$ with min-cut $\rho$ is not solvable, then the dispersion of any scalar linear interpretation for $\Gamma$ is at most $\rho-1$. However, by the max-flow min-cut theorem for the dispersion there exist non-linear interpretations with dispersion arbitrarily close to $\rho$, and hence the highest dispersion may not always be achieved by scalar linear interpretations.

We now significantly strengthen the considerations above by designing a solvable term set for which linear functions have dispersion bounded by a constant, while the min-cut can be arbitrarily large. Let $X = \{x_1,x_2,\ldots,x_k\}$ be a set of variables and consider $k+1$ functions $h_i$ of these variables: $h_i(x_1,x_2,\ldots,x_k) = h_i(X)$. We then define the set $\Gamma$ built on the $k$ variables in $X$ and the function symbols $h_1,h_2,\ldots,h_{k+1}$ together with $f$ of arity $k+1$ and $g_1,g_2,\ldots,g_{k+1}$ of arities all equal to $1$ by
$$
    \Gamma = \left\{ f\Big( g_i\big(h_1(X) \big),h_2(X),h_3(X),\ldots,h_{k+1}(X) \Big) : 1 \leq i \leq k+1 \right\}.
$$

\begin{proposition} \label{prop:gap_linear_solvable}
There exists $n \in \mathbb{N}$ such that for any $A$ with $|A| \geq n$, $\Gamma$ defined above is solvable over $A$, i.e. $\gamma(\Gamma,|A|) = k$. However, any linear interpretation for $\Gamma$ has dispersion at most $2$.
\end{proposition}

In order to prove Proposition~\ref{prop:gap_linear_solvable}, we first consider a term set related to $\Gamma$, where we convert the function symbols $h_1(X)$ into variables.

\begin{lemma} \label{lemma:gap_linear}
Consider the following set of $k+1$ terms built on $k+1$ variables $h_1, h_2,\ldots,h_{k+1}$ and $k+2$ function symbols $f,g_1,g_2,\ldots,g_{k+1}$:
\[
    \Gamma' = \left\{ t_i = f\big( g_i(h_1),h_2,h_3,\ldots,h_{k+1} \big) : 1 \leq i \leq k+1 \right\}.
\]
$\Gamma'$ has min-cut $k+1$, and hence dispersion arbitrarily close to $k+1$. However, if the coding functions are linear, then they have $\Gamma'$-dispersion at most $2$.
\end{lemma}

\begin{IEEEproof}
We first prove that the min-cut of $\Gamma'$ is $k+1$ by constructing $k+1$ vertex-disjoint paths from $\Gamma'_{{\rm var}}$ to $\Gamma'$ in $G_{\Gamma'}$. There is a path $(h_1,g_1(h_1),t_1)$ and $k$ paths $(h_i,t_i)$ for $2 \leq i \leq k+1$; all these paths are clearly vertex-disjoint. By the max-flow min-cut theorem for the dispersion, $\Gamma'$ has dispersion arbitrarily close to $k+1$.

Let $\psi$ be a matrix linear interpretation for $\Gamma'$, i.e. $\bar{g}_i(a_1)= G_ia_1$ for all $1 \leq i \leq k+1$ and $\bar{f}({\bf a}) = \sum_{i=1}^{k+1} F_ia_i$. Consider the induced mapping $\psi$ composed with the permutation $\pi$ of $A^{k+1}$ defined as $\pi({\bf a}) = (a_1,a_2-a_1,\ldots,a_{k+1}-a_1)$. Then we obtain
$$
    \pi \circ \psi({\bf a}) = \left(F_1G_1a_1 + \sum_{i=2}^{k+1} F_i a_i, F_1(G_2-G_1)a_1, \ldots, F_1(G_{k+1}-G_1)a_1 \right),
$$
and $|{\rm image}(\pi \circ \psi)| \leq A^2$. However, we have $|{\rm image}(\pi \circ \psi)| = |{\rm image}(\psi)|$, and hence the interpretation has dispersion at most $2$.
\end{IEEEproof}

We now prove Proposition~\ref{prop:gap_linear_solvable}.

\begin{IEEEproof}
By Theorem \ref{th:dispersion}, there exists $n$ such that for any $A$ with $|A| \geq n$, there is an interpretation $\psi$ for $\Gamma'$ over $A$ with one-to-one dispersion above $k$. Denote the pre-image of ${\rm one}(\psi)$ as $I$, then $|I| \geq |A|^k$. Furthermore, assign the coding functions $\bar{h}_i$ such that the inputs in $A^k$ are mapped to $|A|^k$ elements in $I$. Therefore, the obtained interpretation for $\Gamma$ has perfect dispersion of $k$. On the other hand, since the dispersion of an interpretation for $\Gamma$ is no more than that of an interpretation for $\Gamma'$, any linear interpretation for $\Gamma$ has dispersion at most $2$.
\end{IEEEproof}

The construction above can be easily generalized to obtain the following result.

\begin{corollary} \label{cor:gap_linear}
For any integers $k \geq l \geq 2$, there exists a solvable set $\Gamma$ with dispersion $k$ over all alphabets of sufficient large size where $l$ is the maximal $\Gamma$-dispersion that can be achieved by (matrix) linear coding functions.
\end{corollary}

\subsection{Low-degree solutions} \label{sec:perfect_dispersion}

In Section~\ref{sec:linear_v_non-linear}, we showed that there could be a huge difference between the dispersion achievable by linear coding functions and non-linear coding functions. In this section we show that if the min-cut is achievable by the use of coding functions of low degree (i.e. constant degree independently of the size of the underlying field), then the min-cut is actually achievable by the use of linear coding functions.

More precisely, let $\Gamma = \{t_1,t_2,\ldots,t_r\}$ be a term set built on the variables $\{x_1,x_2,\ldots,x_k\}$ and with min-cut of $\rho$. Let $\psi$ be an interpretation for $\Gamma$ over an alphabet $A$ with perfect $\Gamma$-dispersion of $\rho$. Then {\em decoding functions} for $\psi$ are functions $\bar{d}_1,\bar{d}_2,\ldots,\bar{d}_\rho : A^r \rightarrow A$ such that there exist $i_1,i_2,\ldots,i_\rho$ for which
$$
    (a_{i_1}, a_{i_2}, \ldots, a_{i_\rho}) = \Big(\bar{d}_1 \big(\psi({\bf a}) \big),\bar{d}_2 \big(\psi({\bf a}) \big),\ldots,\bar{d}_\rho \big(\psi({\bf a}) \big) \Big).
$$

\begin{theorem}[Low-degree solutions imply linear solutions] \label{th:low-degree}
Let $\Gamma$ be a term set built on $k$ variables and with min-cut of $\rho$. Assume that there exist coding and decoding functions defined by fixed polynomials $\bar{p}_1,\bar{p}_2,\ldots,\bar{p}_l \in \mathbb{Z}[a_1,a_2,\ldots,a_k]$ with dispersion $\rho$ for arbitrarily large fields $F$ of characteristic $q$. Then there exist scalar linear coding and decoding functions over all sufficiently large fields $F$ of characteristic $q$ that achieve perfect dispersion of $\rho$.
\end{theorem}

\begin{IEEEproof}
Consider a solution $\psi$ based on polynomials of fixed degree. Since the coding and decoding functions are all polynomials of fixed degrees, their compositions are also polynomials of fixed degrees. Suppose $q^k$ is greater than all polynomial degrees and let $L$ be the linear part operator: for any multivariate polynomial $\bar{p}$ over $\mathbb{F}_{q^k}$, $L(\bar{p})$ is the linear part of $\bar{p}$. Then it is easy to check that $L(\bar{p}_1 \circ \bar{p}_2) = L(\bar{p}_1) \circ L(\bar{p}_2)$ if the polynomial $\bar{p}_1 \circ \bar{p}_2$ has degree less than $q^k$. Let us now consider the interpretation $L(\psi)$, defined as taking the linear part of each coding function in $\psi$. The induced mapping of $L(\psi)$ is then equivalent to taking the linear part of the induced mapping of $\psi$. Since $\bar{d}_j \big(\psi({\bf a}) \big) = a_{i_j}$ for all $j$, we have $L(\bar{d}_j) \circ L(\psi) = L(\bar{d}_j \circ \psi) = a_j$, which forms a linear solution with linear decoding functions.
\end{IEEEproof}

We remark that the results in Section~\ref{sec:linear_v_non-linear} imply that conversely, for each characteristic $q$ there exist term sets with solutions which require that at least some of the involved coding functions (including decoding functions) should be given by polynomials of degree at least $|F|$. Example~\ref{example:high_degree_solution} gives a term set with no linear solution, and where a solution is given by polynomials whose degrees depend on the size of the alphabet.

\begin{example} \label{example:high_degree_solution}
Let
$$
    \Gamma = \left\{t_1 = f \big(f(x_1,x_2), f(x_2,x_1) \big), t_2 = g \big(g(x_1,x_2),g(x_2,x_1) \big) \right\}.
$$
This term set has no (scalar) linear solution over fields of characteristic $2$, yet it has non-linear solutions over fields of size divisible by $4$.
\end{example}

\begin{IEEEproof}
Let $\bar{f}$ and $\bar{g}$ be linear functions, i.e. $\bar{f}(a_1,a_2)=\alpha a_1+ \beta a_2$ and $\bar{g}(a_1,a_2) = \gamma a_1+ \delta a_2$, then
$(\bar{t}_1, \bar{t}_2) = \big((\alpha^2+\beta^2)a_1,(\gamma^2+\delta^2)a_1\big)$ does not depend of $a_2$. Therefore, there are no linear solutions.

If $q=2^k$ where $k$ is even, first remark that the polynomial $u^{\sqrt{q}}+u$ has at most $\sqrt{q} < q$ roots in $\mathbb{F}_q$, hence there exists $\tau \in \mathbb{F}_q$ such that $\tau+\tau^{\sqrt{q}} \neq 0$. Then let $f(a_1,a_2)=a_1^{\sqrt{q}}+\tau a_2^{\sqrt{q}}$ and $g(a_1,a_2) = a_1$. We obtain
\begin{eqnarray*}
    (\bar{t}_1,\bar{t}_2) &=& \big( (1+\tau^{\sqrt{q}+1}) a_1 + (\tau+\tau^{\sqrt{q}})a_2,a_1 \big)\\
    (a_1,a_2) &=& \big(\bar{t}_2,(\tau+\tau^{\sqrt{q}})^{-1}(\bar{t}_1-(1+\tau^{\sqrt{q}+1})\bar{t}_2)).
\end{eqnarray*}
Note that the degree of the solution depends on $q$, and hence Theorem~\ref{th:low-degree} does not apply.
\end{IEEEproof}

\section{Case study involving a single coding function} \label{sec:case_study}

The purpose of this section is double. First, we illustrate the different concepts introduced throughout the paper (distributed coding functions, insufficiency of linear coding functions, R\'enyi entropy, etc.). Second, although the results obtained so far are quite tight, we emphasize that for a specific term set, more can usually be said: tighter bounds can be derived, other types of functions can be considered such as linear functions over rings, etc.

Throughout this section, we consider the following term set:
\[
    \Gamma=\{f(x,y), f(x,z), f(w,y), f(w,z)\},
\]
which actually is $\Gamma_k$ introduced in Proposition \ref{prop:no_theorem} for $k=2$. The graph $G_\Gamma$ is given in Figure~\ref{fig:case_study} below. The term set $\Gamma$ can be viewed as an abstraction of a many-to-one cast with four sources and one user, where the intermediate node $f$ corresponds to a relay which only picks up two signals at a time.

\begin{figure}
\begin{center}
    \includegraphics[scale=0.5]{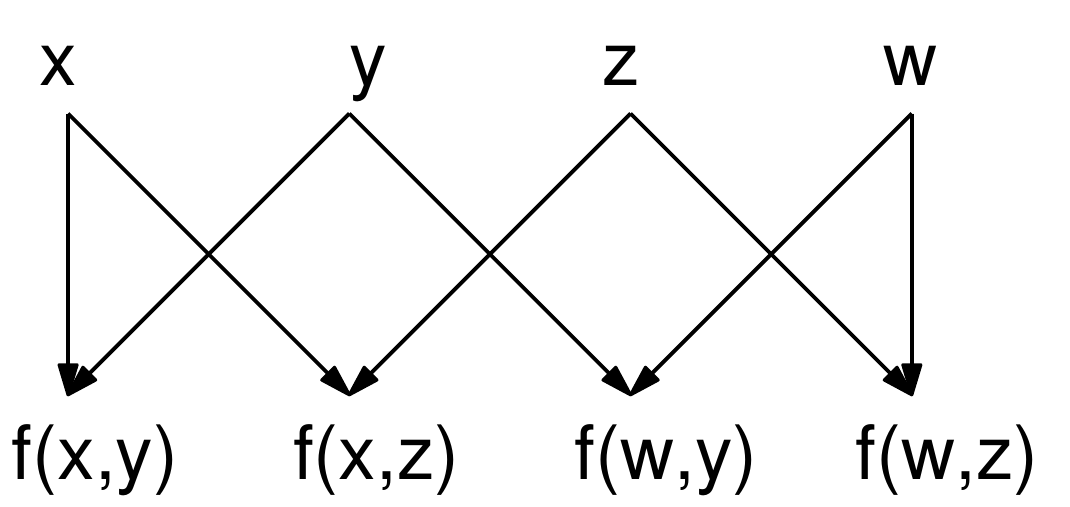}
\end{center}
\caption{The graph $G_\Gamma$.} \label{fig:case_study}
\end{figure}

As seen previously, $\Gamma$ has min-cut $4$ ($\Gamma_{{\rm var}}$ is a term-cut of size $4$; conversely $(x,f(x,y))$, $(z,f(x,z))$, $(y,f(w,y))$, and $(w,f(w,z))$ are $4$ vertex-disjoint paths from $\Gamma_{{\rm var}}$ to $\Gamma$).  According to the max-flow min-cut theorem for the dispersion, for any given $\epsilon>0$ we can select a (coding) function $\bar{f}:A^2 \rightarrow A$ with $\Gamma$-dispersion at least  $4-\epsilon$ for each sufficiently large alphabet $A$. By Proposition~\ref{prop:dispersion_diversified}, the dispersion of the diversified term set is $\gamma_{{\rm one}}(\Gamma^{{\rm div}}, |A|) = \gamma(\Gamma^{{\rm div}}, |A|) = 4$ for all $|A| \geq 2$, and perfect one-to-one dispersion can be achieved by routing for the diversified case. In other words, all demands can be satisfied independently. However, Proposition~\ref{lemma:<4} shows that the whole problem does not have any solution.

\begin{proposition} \label{lemma:<4}
For any alphabet $A$, $\gamma(\Gamma,|A|) < 4$.
In fact we have the tighter bound
\[
    \gamma(\Gamma,|A|) \leq 4 - \log_{|A|}(1 - 2|A|^{-1} + 3|A|^{-2} - |A|^{-3})
\]
\end{proposition}

\begin{IEEEproof}
We prove the tighter bound by refining the argument in the proof of Proposition \ref{prop:no_theorem}. We partition the set of inputs $A^4$ into $4$ parts $C_1,C_2,C_3$, and $C_4$ defined as follows. The set $C_1$ consist of the inputs ${\bf a}$ where $a_1 \neq a_4$ and $a_2 \neq a_3$. This set contains $|A|^2(|A|-1)^2$ points hence the image of $C_1$ contains at most $|A|^2(|A|-1)^2$ points. The set $C_2$ consists of inputs where $a_1=a_4$ and $a_2\neq a_3$, and the set $C_3$ consists of inputs where $a_1\neq a_4$ and $a_2=a_3$. Finally, $C_4$ is the set of inputs where $a_1=a_4$ and $a_2=a_3$. The image of $C_2$ contains at most $|A|(|A|-1)$ points (same for $C_3$) while the image of $C_4$ contains at most $|A|$ points. Thus the range contains at most $|A|^2(|A|-1)^2 + 2 |A|(|A|-1) + |A| = |A|^4-2|A|^3+3|A|^2-|A|$ outputs.
\end{IEEEproof}

We proved in Section~\ref{sec:linear_v_non-linear} that linear maps defined over fields are not optimal when the term set is not solvable. We extend this result below for linear maps defined over rings.

\begin{proposition} \label{lemma:ring}
Assume $A$ is organized as a ring and that $\bar{f}: A^2 \rightarrow A$ is a linear map in the usual sense of algebra, i.e. $\bar{f}(a_1,a_2)=r_1a_1+r_2a_2$ for some $r_1,r_2 \in A$. Then $\bar{f}$ has $\Gamma$-dispersion at most $3$, which is achieved for $r_1 = r_2 = 1$.
\end{proposition}

\begin{IEEEproof}
We first prove the upper bound. Notice that the ring might not be commutative so we do not assume that $r_1r_2 = r_2r_1$. Then
\[
    \psi(a_1,a_2,a_3,a_4) = (r_1 a_1+ r_2 a_2, r_1 a_1 + r_2 a_3, r_1 a_4+r_2 a_2, r_1 a_4+r_2 a_3) \in A^4,
\]
which is uniquely determined by the word $(r_1 a_1+r_2 a_2,  r_2 (a_2-a_3), r_1 a_4+r_2 a_3) \in A^3$. We conclude that the size of the image of $\psi$ is no more than $|A|^3$. This proves the upper bound. For the lower bound let $\bar{f}(a_1,a_2)=a_1+a_2$ and consider the set $C \subseteq A^4$ of inputs with last coordinate equal to $0$; we have
\[
    \psi(a_1,a_2,a_3,0)=(a_1 + a_2, a_1 + a_3, a_2, a_3) \in A^4
\]
and hence the image of $C$ has size $|A|^3$. Thus $\bar{f}$ has a $\Gamma$-dispersion equal to $3$.
\end{IEEEproof}

Similarly, if the alphabet is organized as a group $G$, we define the coding function $\bar{f}_G : G^2 \rightarrow G$ as $\bar{f}_G(\alpha,\beta)=\alpha \beta$ for all $\alpha,\beta \in G$. Then it can be easily shown that $\bar{f}_G$ has $\Gamma$-dispersion of $3$ for any group $G$.

We would like to illustrate the difference between performance measures introduced in Section~\ref{sec:renyi} by considering $A = \mathbb{F}_2$. In this case, there are $2^{2^2} =16$ choices for the coding function $\bar{f}$. However, it can be easily shown that any function is equivalent in terms of the random variable $\psi({\bf a})$ to one of the following four functions: $\bar{f}_0(a_1,a_2) = 0$, $\bar{f}_1(a_1,a_2) = a_1$, $\bar{f}_2(a_1,a_2) = a_1+a_2$, and $\bar{f}_3(a_1,a_2) = a_1a_2$. We easily obtain that $\bar{f}_0$, $\bar{f}_1$, and $\bar{f}_2$ have dispersion $0$, $2$, and $3$, respectively; since they are linear functions, they all have one-to-one dispersion equal to $-\infty$. On the other hand, the behavior of the non-linear function $\bar{f}_3$ is more complex, as its image consists of $10$ elements: the $9$ elements of its one-to-one image together with the all-zero vector, which has $7$ pre-images. Therefore, $\bar{f}_3$ has dispersion $\gamma(\bar{f}_3) = \log_2 10 \equiv 3.32$ and one-to-one dispersion $\gamma_{{\rm one}}(\bar{f}_3) = \log_2 9 \equiv 3.17$. On the other hand, $\bar{f}_3$ has Shannon entropy equal to $H_1(\bar{f}_3) = 4 - \frac{7}{16}\log_2 7 \equiv 2.77$, which is lower than the Shannon entropy of $\bar{f}_2$. Furthermore, $\bar{f}_3$ has min-entropy of $H_\infty(\bar{f}_3) = 4 - \log_2 7 \equiv 1.19$, which is lower than those of $\bar{f}_1$ and $\bar{f}_2$.

\begin{figure}
\begin{center}
    \includegraphics[scale=0.6]{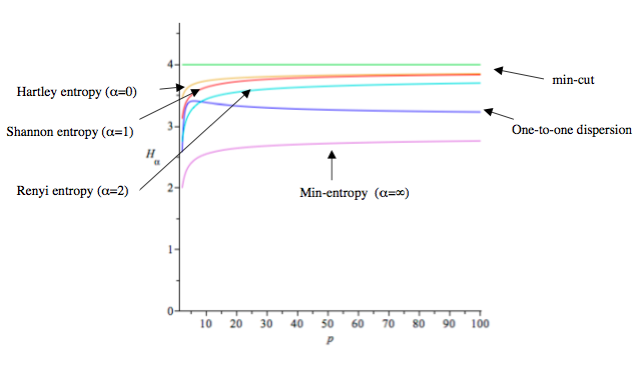}
\end{center}
\caption{Dispersion and R\'enyi entropy of the interpretation $\psi_p$ ($p$ prime)} \label{fig:entropy}
\end{figure}

We now consider the reverse illustration: we fix the coding function, but we change the alphabet on which it is defined. By use of computer calculations, we can show that the coding function $\bar{f}(a_1,a_2)=(a_1-a_2)^2+a_1+a_2$ over the ring $\mathbb{Z}_3$ provides optimal $\Gamma$-dispersion of $\log_3 51 \equiv 3.58$, which attains the upper bound in Proposition \ref{lemma:<4}. The same coding function also provides an optimal one-to-one $\Gamma$-dispersion of $\log_3 36 \equiv 3.26$. In general, we shall denote the interpretation for $\Gamma$ based on the function $\bar{f}$ defined above over $\mathbb{Z}_{|A|}$ as $\psi_{|A|}$. Using an argument based on the Fourier transform \cite{KS10}, one can fully determine the number of elements of the image of that interpretation with a given number of pre-images when the alphabet size is a prime number: $|A| = p$. Thus the entropy $H_\alpha(\psi_p)$ can be determined for all $\alpha$; its $\Gamma$-dispersion and one-to-one $\Gamma$-dispersion are respectively given by
\begin{eqnarray} \nonumber
    \gamma(\psi_p) &=& 4 - \log_p 2 + \log_p(1+p^{-1}-p^{-2}+p^{-3}),\\
    \label{eq:one-to-one_gamma_p}
    \gamma_{\rm one}(\psi_p) &=&  3+ \log_p 3 + 2\log_p (1-p^{-1}).
\end{eqnarray}
The behavior of the R\'enyi entropy for different values of $\alpha$ and the one-to-one dispersion of $\psi_p$ is displayed in Figure \ref{fig:entropy}. On the other hand, when $p$ tends to infinity, then the R\'enyi entropy actually reaches the upper bound from Proposition \ref{prop:no_theorem}. A more detailed argument is given in the next paragraph. On the other hand, by (\ref{eq:one-to-one_gamma_p}) the one-to-one dispersion of $\psi_p$ tends to $3$ on the primes $p$, which is below the min-cut. Thus, the interpretation $\psi_p$ is asymptotically optimal in terms of R\'enyi entropy but not in terms of one-to-one dispersion. This illustrates the fundamental difference between the one-to-one dispersion and the different entropy measures.

Using an idea by Keevash and Sisask based on Fourier analysis \cite{KS10}, it is possible to show that the image of $\psi_p$ for each prime number $p$ is partitioned into $4$ sets $S_1, S_2, S_p, S_{3p-2}$ where:
\begin{itemize}
    \item $S_1$ contains $3p(p-1)^2$ points with exactly one preimage

    \item $S_2$ contains $p(p-1)^2(p-3)/2$ points with exactly two preimages

    \item $S_{p-1}$ contains $2p(p-1)$ points with exactly $p-1$ preimages

    \item $S_{3p-2}$ contains $p$ points with exactly $3p-2$ preimages.
\end{itemize}
Therefore, the size of the image is $\frac{p}{2} (p^3+p^2-p+1)$. Moreover, $\psi_p$ has R\'enyi entropy of order $\alpha \neq 1$ given by
\[
    H_\alpha(\psi_p) = \frac{1}{1-\alpha}\log_p\big( 3(p-1)^2 + (p-1)^2(p-3) 2^{\alpha-1} + 2(p-1)(p-1)^{\alpha} + (3p-2)^{\alpha} \big) + \frac{1-4\alpha}{1-\alpha}
\]
The limit of the R\'enyi entropy (including the case where $\alpha=1$) when the size of the alphabet tends to infinity can be shown to be given by
\[
    \lim_{p \rightarrow \infty} H_\alpha(\psi_p) = \left \{ \begin{array}{ll}
    4 & \mbox{if }  0  \leq \alpha \leq 2 \\
    \frac{3\alpha-2}{\alpha-1} &  \mbox{if } 2 \leq \alpha < \infty  \\
    3 & \mbox{if } \alpha = \infty
       \end{array}\right.
\]
This shows that the bound $\frac{(2k-1)\alpha-k}{\alpha -1}$ is a matching upper and lower bound for $k=2$.

Computer calculations show that the dispersion and one-to-one dispersion of $\psi_p$ behave in a slightly irregular fashion on composite numbers. We notice for instance that the one-to-one dispersion is surprisingly large for prime powers, as seen in Figure \ref{fig:dispersionodd}.

\section{Network Coding solvability and term sets} \label{sec:NC_terms}

\subsection{Converting a multi-user channel into a single-user channel} \label{sec:multi}

We now consider multi-user communications; our approach is to convert multi-user channels into single-user channels. To each user $1 \leq j \leq m$, we associate a term set $\Gamma_j$.  Each choice of coding functions determines a dispersion of $(\Gamma_1,\Gamma_2,\ldots,\Gamma_m)$ which is an array of dispersions. Let $X = \{x_1,x_2,\ldots,x_k\}$ be the set of variables on which all terms are built. For any $1 \leq j \leq m$, let $X^j = \{x_1^j,x_2^j,\ldots,x_k^j\}$ be a new set of variables, and let $\Gamma^j$ be the term set obtained from $\Gamma_j$ by replacing the variable $x_i$ by $x^j_i$ for all $1 \leq i \leq k$. It is clear that any interpretation for $\Gamma_j$ can be viewed as an interpretation for $\Gamma^j$, and that its $\Gamma_j$-dispersion (also one-to-one dispersion and R\'enyi entropy) is equal to its $\Gamma^j$-dispersion.

Consider now $\bar{\Gamma} := \bigcup_{j=1}^m \Gamma^j$, where the union is disjoint and with min-cut equal to $\rho = \sum_{j=1}^m \rho_j$. In fact the graphs $G_{\Gamma^j}$ are components of $G_{\bar \Gamma}$, which shows that for any interpretation $\psi$ for $\bar{\Gamma}$, we have $\gamma(\psi) = \sum_{j=1}^m \gamma(\psi^j)$, where $\psi^j$ is the corresponding interpretation for $\Gamma^j$. We then conclude that there exists an interpretation for $\bar{\Gamma}$ over $A$ with dispersion $\rho$ if and only if there exist $m$ interpretations for $\Gamma_j$ with dispersions $\rho_j$ for all $1 \le j \le m$. Moreover, applying the max-flow min-cut theorems for the term set $\bar{\Gamma}$, we obtain the multi-user max-flow min-cut theorem below.

\begin{theorem}[Multi-user max-flow min-cut theorem] \label{th:multi-user}
Let $\Gamma_1,\Gamma_2,\ldots,\Gamma_m$ be term sets with respective min-cuts $\rho_1,\rho_2,\ldots,\rho_m$. Then for any $\epsilon > 0$, there exists $n_0$ such that for all $A$ with $|A| \geq n_0$, $\gamma_{\rm one}(\Gamma_j,|A|) \geq \rho_j - \epsilon$ for all $1 \leq j \leq m$. Also, for any $0 \leq \alpha < 1$, there exists $n_\alpha$ such that $H_\alpha(\Gamma_j,|A|) \geq \rho_j - \epsilon$ for all $|A| \geq n_\alpha$.
\end{theorem}

\subsection{Multi-user communication problems and term sets} \label{sec:NC_solvability}

We now apply the multi-user max-flow min-cut theorem to communication problems. A many-to-many cast is defined as follows.

\begin{definition} \label{def:NC}
A multi-user communication problem instance (also referred to as a {\em many-to-many cast}) is a tuple $(V,E,S,U,A)$, where
\begin{itemize}
    \item $G = (V,E)$ is an acyclic directed graph, where the vertices $v_1,v_2,\ldots,v_{|V|}$ of $V$ are sorted such that $(v_i,v_j) \in E$ only if $i<j$.

    \item $S= \{s_1,s_2,\ldots,s_k\} \subseteq V$ is the set of sources, which are nodes with in-degree $0$. Without loss of generality, $s_i = v_i$ for $1 \leq i \leq k$.

    \item $U = \{r_1,r_2,\ldots,r_m\} \subseteq V$ is the set of users (receivers), which are nodes with out-degree $0$. Without loss, $r_j = v_{|V|-m+j}$ for $1 \leq j \leq m$.

    \item $A$ is an alphabet of size $|A| \geq 2$. Each source $s_i$ sends a distinct message $a_i \in A$, and each user $r_j$ requests $\{a_1,a_2,\ldots,a_k\}$. 
\end{itemize}
Each vertex $v_k \in V \backslash S$ can manipulate the data it receives and transmit a function $f_k$ of its inputs onto all its out-edges.
\end{definition}

A communication problem can be equivalently defined with terms. Each user obtains a term built on the variables sent by the sources, where the function symbols represent the operations made by the intermediate nodes in $V \backslash (S \cup U)$. More formally, to each source $s_i$ we associate the variable $x_i$ and we denote $X = \{x_1,x_2,\ldots,x_k\}$. Each user $r_j$ requests all the variables in $X$. We then associate the function symbol $f_l$ to all intermediate nodes $v_l \in V \backslash (S \cup U)$ and each vertex $v_l$ is recursively assigned the term $u_l = f_l(u_{l_1},u_{l_2},\ldots,u_{l_d})$, where $\{v_{l_1},v_{l_2},\ldots,v_{l_d}\}$ is the in-neighborhood of $v_l$. We denote the in-neighborhood of the user $r_j$ as $\{v_{j,1},v_{j,2},\ldots,v_{j,p_j}\}$. Note that using this notation, it is possible that $v_{j,i} = v_{j',i'}$ for distinct $j,j'$ and $i,i'$. However, this will not affect our definitions below. We finally associate the term $t_{j,i}$ to the vertex $v_{j,i}$.

A solution for the many-to-many cast instance is a choice of the functions at the intermediate nodes such that all the users' demands can be satisfied at the same time. Using terms, user $j$ is satisfied if and only if it can recover $X$ from the term set $\Gamma_j = \{t_{j,i}\}_{i=1}^{p_j}$ obtained from its in-neighborhood. This implies that the term set $\Gamma_j$ must have dispersion of $k$. In order to take into account the fact that all demands have to be satisfied at the same time, we use the term set $\bar{\Gamma}$ defined by diversifying variables. By our construction and the remarks on $\bar{\Gamma}$ made above, we obtain the following result.

\begin{proposition} \label{prop:NC}
A many-to-many cast instance is solvable over $A$ if and only if $\gamma_{{\rm one}}(\Gamma_j,|A|) = \gamma(\Gamma_j,|A|) = k$ for all $1 \leq j \leq m$ or equivalently,
\begin{equation} \label{eq:dispersion_NC} \nonumber
    \gamma_{{\rm one}}(\bar{\Gamma},|A|) = \gamma(\bar{\Gamma},|A|) = km.
\end{equation}
\end{proposition}

The multi-user max-flow min-cut theorem then shows that if the min-cut between the sources and each user is equal to $k$, then the multi-user instance is asymptotically solvable. Note that the equivalence in Proposition \ref{prop:NC} together with Theorem \ref{th:multi-user} indicate that if each user can be asymptotically satisfied individually, then all the users' demands can asymptotically be satisfied simultaneously.

The following two examples are consequences of Proposition \ref{prop:NC}.

\begin{example} \label{ex:butterfly}
In \cite{YZ99}, Yeung and Zhang made a simple observation that lay the foundation for network coding. The authors considered a situation where two users communicate via a satellite. User $X$ wants to send a message $x\in A$  to $Y$, while $Y$ at the same time wants to send a message $y \in A$ to $X$.
\end{example}

\begin{figure}
\begin{center}
    \includegraphics[scale=0.45]{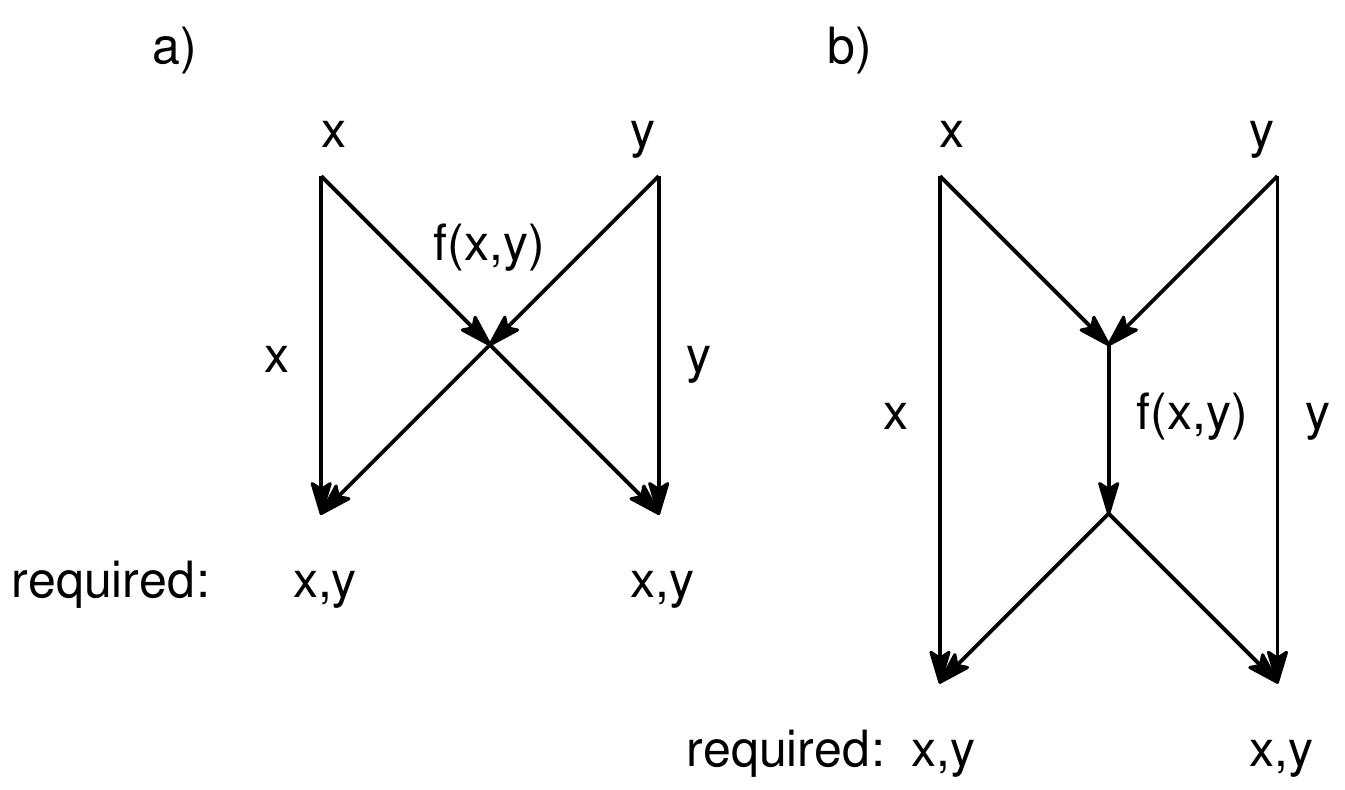}
\end{center}
\caption{Butterfly network.} \label{fig:butterfly}
\end{figure}

The satellite communication problem is equivalent to the communication problem in Figure~\ref{fig:butterfly}a), referred to as the butterfly network. In Figure~\ref{fig:butterfly}a), the function symbols are affected to the vertices, accordingly to Definition~\ref{def:NC}; however the problem is sometimes represented in the literature by Figure~\ref{fig:butterfly}b), where the function symbols are affected to the edges. A solution over an alphabet $A$ is a function $\bar{f}:A^2  \rightarrow A$ with the property that there exist decoding functions $\bar{d}_1, \bar{d}_2:A^2 \rightarrow A$ such that $b=\bar{d}_1\big(b,\bar{f}(a,b)\big)$ and  $a=\bar{d}_2\big(a,\bar{f}(a,b)\big)$ for any $(a,b) \in A^2$.

By Proposition~\ref{prop:NC}, the satellite problem (and the butterfly communication problem) is mathematically equivalent to the following problem on terms.

\begin{problem}[Equivalent to the satellite communication problem] \label{pr:satellite}
Construct a function $\bar{f}:A^2 \rightarrow A$ with $\Gamma$-dispersion equal to $4$, where $\Gamma:=\{x_1, f(x_1,x_2), f(x_3,x_4), x_4\}.$
\end{problem}

\begin{figure}
\begin{center}
    \includegraphics[scale=0.60]{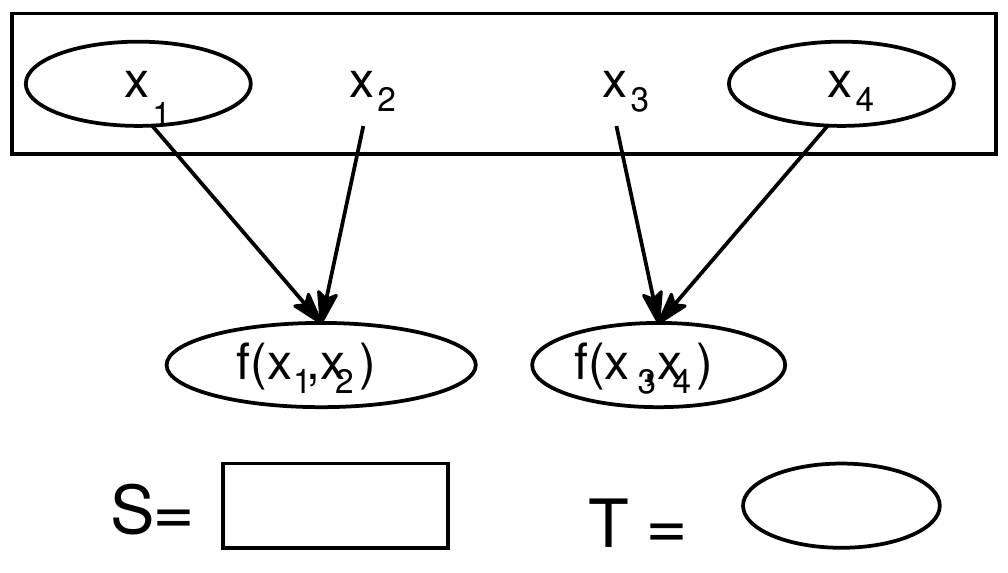}
\end{center}
\caption{Graph $G_\Gamma$ for the butterfly network.} \label{fig:G_Gamma_butterfly}
\end{figure}

This term set corresponds to the graph with vertex set
\[
    V=\Gamma_{\rm sub}=\{x_1,x_2,x_3,x_4,f(x_1,x_2), f(x_3,x_4)\},
\]
source set $S=\{x_1,x_2,x_3,x_4\}$ and target set $T=\Gamma$. The edge set $E \subseteq V \times V$ is given by
\[
    E = \left\{ \big(x_1,f(x_1,x_2) \big), \big(x_2,f(x_1,x_2) \big), \big(x_3,f(x_3,x_4) \big), \big(x_4,f(x_3,x_4) \big) \right\}.
\]
The graph $G_{\Gamma}$ is displayed in Figure~\ref{fig:G_Gamma_butterfly}.

Problem~\ref{pr:satellite} can be viewed as a communication problem between a single source that transmits a message $(a_1,a_2,a_3,a_4) \in A^4$ and a user who receives a message of the form $\big(a_1, \bar{f}(a_1,a_2), \bar{f}(a_3,a_4), a_4 \big) \in A^4$.

\begin{example}[Distributed storage] \label{example:storage}
Assume that we want to store two messages $x,y \in A$ at four locations. The messages $x$ and $y$ are stored at two of the locations. At the two remaining locations two messages $f(x,y) \in A$ and $g(x,y) \in A$ are stored. The problem is to select the coding functions $f:A^2 \rightarrow A$ and $g:A^2 \rightarrow A$ such that it is always is possible to reconstruct $x$ and $y$ from accessing only two of the four locations.
\end{example}

This type of problem has already been studied in the literature \cite{DGWWR10} as part of network coding, as well as an application of error correcting codes. The actual problem can be shown to be equivalent to the existence of two orthogonal Latin squares of order $|A|$ \cite{RA05}. This problem was first posed by Euler around 1780 and was eventually completely solved in 1960, where it was shown in \cite{BSP60} that there exist orthogonal Latin squares of any order except of order $2$ and order $6$.
The distributed storage problem is mathematically equivalent to the following problem.

\begin{problem}[Equivalent to the distributed storage problem] \label{pr:storage}
Construct two functions $\bar{f},\bar{g}:A^2 \rightarrow A$ with $\Gamma$-dispersion equal to $10$, where
\[
    \Gamma:= \{ x_1, f(x_1,y_1), y_2, f(x_2,y_2),  x_3, g(x_3,y_3), y_4, g(x_4,y_4), f(x_5,y_5),  g(x_5,y_5)\}.
\]
\end{problem}

\begin{IEEEproof}
We can view the storage problem as a many-to-many cast instance with two sources sending $x$ and $y$ and six users obtaining $\{x,y\}$, $\{x, f(x,y)\}$, $\{y, f(x,y)\}$, $\{x, g(x,y)\}$, $\{y, g(x,y)\}$, and $\{f(x,y), g(x,y)\}$ respectively. The demands of the first user are trivially satisfied, so only the last five need to be considered. Applying the transformations above and Proposition~\ref{prop:NC}, we obtain the desired term set.
\end{IEEEproof}

Problem~\ref{pr:storage} can be viewed as a communication problem between a single source that transmits a message ${\bf a} \in A^{10}$ and a user who receives a message $\psi({\bf a}) \in A^{10}$.

We can give a network coding interpretation of the diversified term set. When the term set is diversified, then we can select a different solution for each user independently. Therefore, we have full dispersion in the diversified case if and only if each request can be satisfied individually.

\begin{example} \label{example:butterfly_diversified}
Consider the term set $\Gamma=\{x_1,f(x_1,x_2),x_4,f(x_3,x_4)\}$ corresponding to the satellite communication problem. The diversified term set can be written as $\Gamma^{{\rm div}} = \{x_1,f(x_1,x_2),x_4,g(x_3,x_4)\}$. Since the coding functions $f$ and $g$ are not required to be identical, it is easier to find coding functions that achieve the maximal dispersion of $4$ in the case of $\Gamma^{{\rm div}}$. One possible choice of coding functions that achieve dispersion of $4$ is to let $f(x_1,x_2)=x_2$ and let $g(x_3,x_4)=x_3$. However, this choice does not correspond to any real-life situation for the original satellite communication problem. In fact, diversifying the term set is equivalent to considering the case where the satellite has access to two independent channels to the stations on which it can send a different message (one constructed by the function $f$, the other by the function $g$).
\end{example}

The max-flow min-cut theorem for the dispersion then shows that if all demands can be satisfied individually, then for all large enough alphabets, all demands can be ``nearly'' satisfied at once. Theorems~\ref{th:dispersion} and~\ref{th:dispersion_refined} quantify this statement in terms of a small loss in one-to-one dispersion. However, this small loss may be critical when network coding is considered. Indeed let $\Gamma$ be the term set associated to a given multi-user communication problem, and let $\psi$ be the induced function of an interpretation for $\Gamma$. Then, the one-to-one pre-image of $\psi$ may contain very few points $({\bf a},{\bf a},\ldots,{\bf a}) \in A^{km}$, where ${\bf a} \in A^k$. However, only these points make sense for the multi-user communication, as the variable $x_i^j$ is merely an artificial variable representing $x_i$ for all $j$. Consider the butterfly network for example. Then all the points in the one-to-one pre-image of dynamic routing satisfy $x_1 \neq x_3$, which does not correspond to any real-life situation for the satellite communication problem.

\section{Dynamic networks} \label{sec:extension}

\subsection{Theorem for multi-user dynamic networks} \label{sec:extension_intro}

In the analysis of dynamic communication networks it is natural to take into account that networks change over time. Potential network changes including link failures, point failures and noisy channels can be modeled by ideas vaguely akin to Kripke's {\em possible world semantics} from logic and philosophy \cite{K59}. The idea is to consider a collection of {\em possible worlds} which each could become {\em the actual world} as time progresses.

A world is not only a representation of the network at a given time, but an expansion over a number of time slots. This is a generalization of the butterfly network, which can be viewed as an expansion over time of the satellite communication problem. As the model is discrete there are only a finite a number $|W|$ of possible worlds. We can think of each node in a world as a network node at a certain time slot.  A point in each world thus represents a node at a given time slot, and as such there is a link from each node to its successor in time.  This link is not a communication link, but represents the data transformation at the node in the two time slots. Communications might be instantaneous  i.e. connect different nodes in the same time slot.  The resulting network is acyclic. Formally we define

\begin{definition}[Dynamic network] \label{defi: Dynamic}
Let $U,W$, and $T$ be finite sets.  A {\em dynamic multiuser network} is a collection $\tilde{\Gamma} := \{\Gamma_{u,w,t}: u \in U, w \in W, t \in T \}$ of term sets indexed by a user/receiver $u \in U$, a world $w \in W$, and a time-slot $t \in T$. Each $\Gamma_{u,w,t}$ consists of a collection of terms as well as a collection of variables required.  If each $\Gamma_{u,w,t}$ require all the variables the network is
a {\em dynamic many-to-many cast (or multi-cast) network} 

\end{definition}

Assume all coding functions that occur in terms in $\tilde{\Gamma}$ have been given interpretations. Then we can associate a dispersion to each set term set $\Gamma_{u,w,t}$. For each set $\Gamma_{u,w,t}$ we associate a variable $\gamma_{u,w,t}$ which for each choice of coding functions denotes the dispersion of the term set $\Gamma_{u,w,t}$. Typically the same function symbol might occur in terms sets in multiple worlds. In general a coding function might be a good choice for some of the worlds, while it might be a bad choice for other worlds.

In order to take into account the fact that some worlds may be more likely than others, we consider the general case where each user $u \in U$ is assigned a {\em utility demand} $D_u$. This is of the form $D_u \equiv F_u(\gamma_u) > {\rm dem}_u$, where ${\rm dem}_u$ is a real number and the utility function $F_u$ is a real-valued, non-decreasing, and continuous function in the variables $\gamma_u = \{\gamma_{u,w,t} : w  \in W, t \in T\}$ representing the received dispersions.
We say that the demand $D_u$ is satisfied {\em locally} if it can be satisfied when all other user demands are disregarded. Conversely, we say that $D_1, D_2,\ldots,D_m$ are satisfied {\em globally} if they can all be satisfied by the same interpretation.

\begin{theorem}[Dynamic multi-user theorem] \label{th:demands}
In a dynamic many-to-many cast where the demand $D_u$ of each user $1 \leq u \leq m$ can be satisfied locally, the demands $D_1,D_2,\ldots,D_m$ can be satisfied globally. The same holds for R\'enyi entropy demands with $\alpha < 1$ and one-to-one dispersion demands.
\end{theorem}

\begin{IEEEproof}
Suppose that each $F_u(\gamma_u)>{\rm dem}_u$ with $1\leq u \leq m$ can be achieved locally. If we select $\delta >0$ such that $\delta < {\rm min}_u \{F_u(\gamma_u) - {\rm dem}_u\}$, then in fact $F_u(\gamma_u) > {\rm dem}_u+\delta$ with $1\leq u \leq m$  can be achieved locally. Assume that  $F_u(\gamma_u) > {\rm dem}_u+\delta$ is achieved (locally) by the dispersions $\gamma_u = \{\gamma_{u,w,t} : w \in W, t \in T\}$. Since $F_u$ is continuous, there exists $\epsilon > 0$ such that if the dispersion $\gamma'_u$ of $\{\Gamma_{u,w,t} : w \in W, t \in T\}$ has $| \gamma'_{u,w,t} - \gamma_{u,w,t} | < \epsilon$, then $|F_u(\gamma'_u)- F_u(\gamma_u)|<\delta$. According to the multi-user max-flow min-cut theorem, for each $\epsilon>0$ there exists an interpretation (over a sufficiently large alphabet) which globally achieves the dispersions $\gamma_{u,w,t} - \epsilon$ , $u \in U, w \in W, t \in T$. Thus there exist coding functions such that $F_u(\gamma_u) > {\rm dem}_u$ for $1 \leq u \leq m$.
\end{IEEEproof}

We remark that the demands could also be expressed as $F_u(\gamma_u) \geq {\rm dem}_u$. In that case, the dynamic multi-user theorem indicates that the demands can asymptotically be achieved globally if they can be asymptotically achieved locally. Example \ref{example:utility} below shows how the utility function can cover a broad family of performance measures.

\begin{example} \label{example:utility}
Let $\tilde{\Gamma}$ consist of term sets $\Gamma_{u,w,t}$ where $u \in U, w \in W$, and $t \in T$. Assume world $w$ occurs with probability $p_w$ and the user $u$ has assigned a weight $\omega_t$ proportional to the utility of the dispersion achieved in time slot $t$. Then the utility for each user $u \in U$ can be defined as
\[
  F_u(\gamma_u) = \sum_{w\in W, t \in T} p_w \omega_t  \gamma_{u,w,t}
\]
which is a continuous function in the variables $\gamma_u$. Asymptotically, the maximal utility achievable for user $u$ is given by
\[
  \sum_{w\in W, t \in T}  p_w \omega_t \rho_{u,w,t}
\]
where $\rho_{u,w,t}$ denotes the min-cut of $\Gamma_{u,w,t}$.
\end{example}

The example above can be viewed as an asymptotic generalization of the network coding theorem, which only considers one time slot and one possible world. A proper generalization of that result can be obtained by considering linear coding functions over a a diversified term set for a multi-user communication problem.

\subsection{Clairvoyance and term equations} \label{sec:clairvoyance}

In a dynamic network, various unpredictable network changes (e.g. link failures) might happen in various of the possible worlds. A realistic choice of coding functions cannot look into the future and take into account which link might fail during transmission. Nonetheless, we define clairvoyance as the case where each node ``knows'' in which of the possible worlds the network is.  We can define the formally as follows:

\begin{definition}[Clairvoyant coding] \label{defi: Clairvoyant}
The clairvoyant version $\tilde{\Gamma}^{\rm clair}$ of $\tilde{\Gamma}$ is defined as the collection of term sets where function symbols have been diversified so  function symbols in different worlds are distinct, e.g. each function symbol the occurs in term sets with index $w \in W$ is assigned an (additional) index $w$.
An assignment of functions symbols to $\tilde{\Gamma}^{\rm clair}$ is said to consist of {\em clairvoyant coding functions} for $\tilde{\Gamma}$.
\end{definition}

As another application of the multi-user dispersion theorem we obtain that clairvoyance does not improve the performance of a network in terms of dispersion. The proof of Proposition \ref{prop:clairvoyance} below is based on arguments similar to those used to prove the theorem for the dispersion.

\begin{proposition}[Clairvoyance does not increase dispersion] \label{prop:clairvoyance}
If in a dynamic many-to-many network the users demands $D_1,D_2,\ldots,D_r$ can be satisfied in $\tilde{\Gamma}^{\rm clair}$ then they can be satisfied in $\tilde{\Gamma}$.
\end{proposition}

It should be noticed that this result for the dispersion and the R\'enyi entropy is very much in the spirit of diversity coding and random linear network coding and is thus not surprising. Indeed, diversity coding intuitively deals with link failures and noisy channels by mixing the inputs and transmitting a large number of independent messages. Clairvoyance is thus rendered useless. However, our result about the one-to-one dispersion is remarkable, for--as seen in Section \ref{sec:entropy_model}--a high one-to-one dispersion involves a controlled non-linear mixing, which contradicts the philosophy of diversity coding.

We finish this section by revealing that a model based on term equations can take into account the fact that each user not only requires a high dispersion, but also a certain number of specific messages. Recall from Section \ref{sec:linear} that if a user receives the terms $\{t_1,t_2,\ldots,t_r\}$, we associate the decoding functions $\bar{d}_1,\bar{d}_2,\ldots, \bar{d}_s$ where we require that $\bar{d}_i(\psi({\bf a})) = a_i$ for all $1 \leq i \leq s$. This can be more succinctly expressed in the term equation $\tau_i = x_i$, where $\tau$ is the term defined as $\tau_i = d_i(t_1,t_2,\ldots,t_r)$.

In the case of message demands, then clairvoyance can clearly make a difference, as seen in Example \ref{ex:clairvoyance} below.

\begin{example} \label{ex:clairvoyance}
Consider the dynamic network depicted in Figure \ref{fig:clairA}, where one of two links is always contaminated with pure noise. Remark that this is not the butterfly network in Figure \ref{fig:butterfly}. Without using clairvoyance, the message demands of the destinations can be expressed as term equations
\begin{eqnarray*}
    d_1({\rm noise}_1,f(x,y)) &=& x\\
    d_2(y,f(x,y)) &=& y\\
    d_3(x,f(x,y)) &=& x\\
    d_4({\rm noise}_2,f(x,y)) &=& y.
\end{eqnarray*}
Here we assumed that the decoding functions can distinguish  the messages $x$ and $y$ from noise, which is why we can apply different decoding functions to each of the four potential decoding situations.
It is clear that the message demands of both users cannot be satisfied globally without clairvoyance. On the other hand, using clairvoyance, the problem is turned into the following set of term equations:
\begin{eqnarray*}
    d_1({\rm noise}_1,f_1(x,y)) &=& x\\
    d_2(y,f_1(x,y)) &=& y\\
    d_3(x,f_2(x,y)) &=& x\\
    d_4({\rm noise}_2,f_2(x,y)) &=& y.
\end{eqnarray*}
Thus, letting $\bar{f}_1(x,y) = x$ for world $1$ and $\bar{f}_2(x,y) = y$ for world $2$ solves the communication problem by an appropriate choice of decoding functions $\bar{d}_1,\bar{d}_2,\bar{d}_3$, and $\bar{d}_4$.

\begin{figure}
\begin{center}
    \includegraphics[scale=0.40]{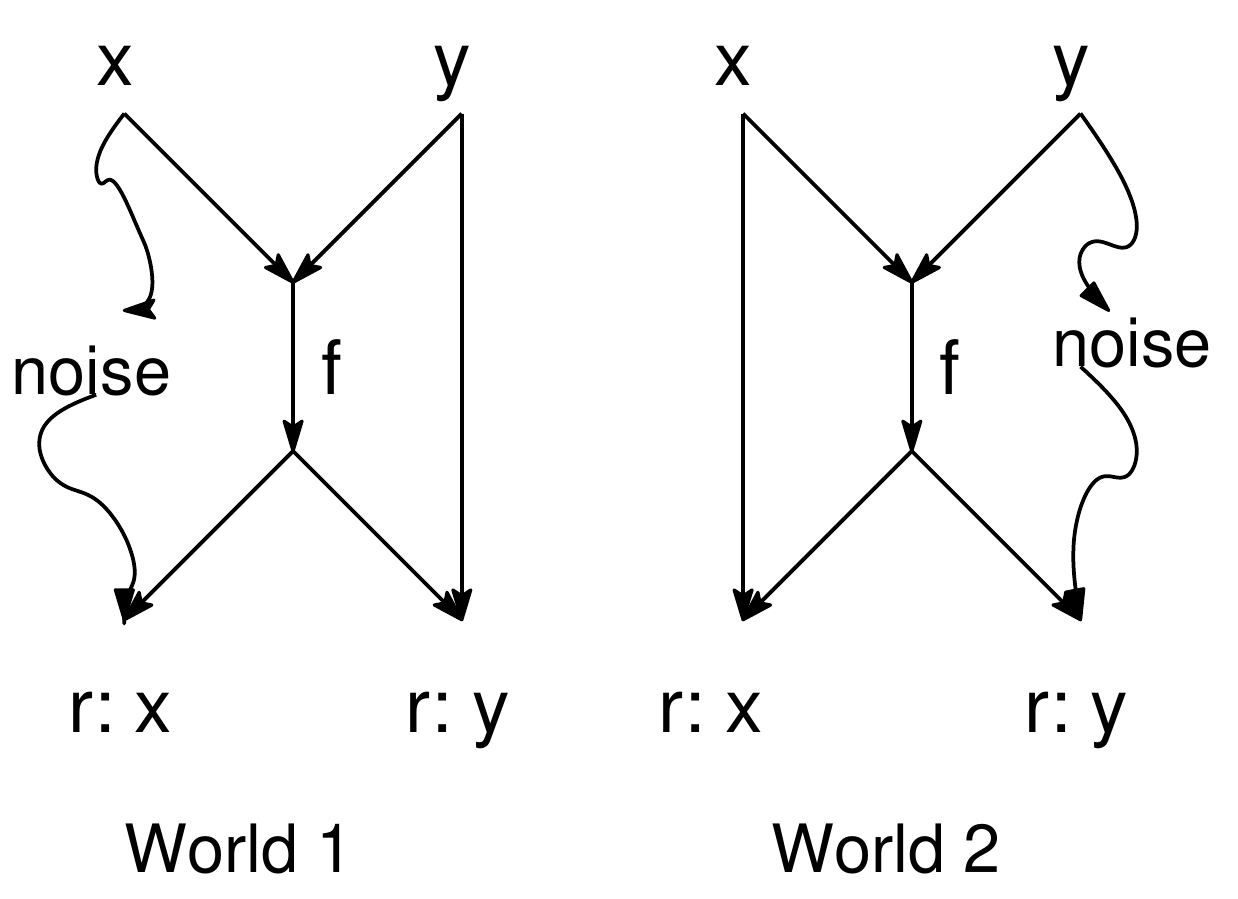}
\end{center}
\caption{Dynamic network with message demands.}\label{fig:clairA}
\end{figure}
\end{example}

In brief, clairvoyance does not help in many-to-many casting where the users have dispersion demands.  But not surprisingly in general--when different users have different message demands--clairvoyance can greatly increase the performance of the network.

\section{Conclusion} \label{sec:conclusion}

There is an extensive literature for dealing with the logistics and scheduling in traditional commodity networks. The theories are very diverse ranging from linear programming, algorithms for transport of ``discrete'' goods, game theory, traffic flow theory, network exchange theory, economic network theory, packet switching, and queuing theory. It is less obvious that transport of digital information has a cost and that scheduling traffic of data is beneficial. Maybe this is why it historically was very late that people have begun (mainly in the field of Network Coding) to develop theories that cover transport of digital information in communication networks.  

In this paper we developed a general theory for transport of digital information in relay networks.  To summarise, we

\begin{itemize}

\item considered relay networks and explained how such networks can be used to model dynamic communication networks,

\item introduced a formalism (the term model) that made it possible to handle communication problems in a graph-free approach,

\item showed that the asymptotic throughput of any single-sender single-receiver relay network is given my its minimal term-cut,

\item showed that any question about solvability of a given dynamic multi-user communication problem,  can be restated as a question about solvability of a single-sender single-receiver problem in a specific relay network. 

\end{itemize}

\bibliographystyle{IEEEtran}
\bibliography{g}

\end{document}